\documentclass{aa}
\usepackage[varg]{txfonts}
\usepackage{float}
\usepackage[caption = false]{subfig}
\usepackage{caption}
\usepackage{amsmath}
\usepackage{amssymb}
\usepackage{graphicx}
\usepackage{array}
\usepackage{pgfplotstable}
\usepackage[utf8]{inputenc}
\usepackage{booktabs}
\usepackage{marvosym}
\usepackage{lipsum}
\usepackage{times}
\usepackage{multirow}
\usepackage{rotating}
\usepackage{longtable}
\usepackage{lscape}
\usepackage{epstopdf}
\usepackage{hyperref}
\usepackage{mathrsfs}
\usepackage{etoolbox}
\makeatletter
\patchcmd\@combinedblfloats{\box\@outputbox}{\unvbox\@outputbox}{}{%
   \errmessage{\noexpand\@combinedblfloats could not be patched}%
}%
 \makeatother

\title{The VLA-COSMOS 3 GHz Large Project: Star formation properties and radio luminosity functions of AGN with moderate-to-high radiative luminosities out to $z\sim6$}
\titlerunning{Star formation properties and radio LFs of HLAGN out to $z\sim6$}
\author{L. Ceraj\inst{1}
	\thanks{\emph{lceraj@phy.hr}}
	\and V. {Smol\v{c}i\'{c}} \inst{1}
	\and I. Delvecchio \inst{1}
	\and M. Novak \inst{1,2}
	\and G. Zamorani \inst{3}
	\and J. Delhaize \inst{1,6}
    \and E.~Schinnerer\inst{2}
    \and E. Vardoulaki\inst{4}
    \and N. Herrera Ruiz\inst{5}
	}
\institute{University of Zagreb, Physics Department, Bijeni\v{c}ka cesta 32, %1
	10002 Zagreb, Croatia.
  \and  Max Planck Institut f\"ur Astronomie, K\"onigstuhl 17, 69117
Heidelberg, Germany.
  \and  INAF - Osservatorio di Astrofisica e Scienza dello Spazio - Bologna, Via Piero Gobetti
93/3, I-40129 Bologna, Italy.
	\and Argelander-Institut f\"ur Astronomie, Universit\"at Bonn, Auf dem H\"ugel 71, D-53121 Bonn, Germany.
	\and Astronomisches Institut, Ruhr-Universit\"at Bochum, Universit\"atsstrasse 150, 44801 Bochum, Germany.
    \and Department of Astronomy, University of Cape Town, Private Bag X3, Rondebosch 7701, South Africa
	}

\begin{document}

%\date{Received 2 November 2016 Accepted 7 January 2016}		

\abstract{We study a sample of $1,604$ moderate-to-high radiative luminosity active galactic nuclei (HLAGN) selected at 3 GHz within the VLA-COSMOS 3 GHz Large Project. 
These were classified by combining multiple AGN diagnostics: X-ray data, mid-infrared data and broad-band spectral energy distribution fitting. 
We decompose the total radio 1.4 GHz luminosity ($\mathrm{L_{1.4\ GHz, TOT}}$) into the emission originating from star formation and AGN activity by measuring the excess in $\mathrm{L_{1.4\ GHz, TOT}}$ relative to the infrared-radio correlation of star-forming galaxies. 
To quantify the excess, for each source we calculate the AGN fraction ($\mathrm{f_{AGN}}$) defined as the fractional contribution of AGN activity to $\mathrm{L_{1.4\ GHz, TOT}}$.
The majority of the HLAGN, $(68.0\pm1.5)\%$, are dominated by star-forming processes ($\mathrm{f_{AGN}\leq0.5}$), while $(32.0\pm1.5)\%$ are dominated by AGN-related radio emission ($\mathrm{0.5<f_{AGN}\leq1}$).
We use the AGN-related 1.4 GHz emission to derive the 1.4 GHz AGN luminosity functions of HLAGN. 
By assuming pure density and pure luminosity evolution models we constrain their cosmic evolution out to $\mathrm{z\sim6}$, finding $\mathrm{\Phi^* (z) \propto (1+z)^{(2.64\pm 0.10) + (-0.61\pm 0.04) z}}$ and $\mathrm{L^* (z) \propto (1+z)^{ (3.97\pm 0.15) + (-0.92\pm 0.06) z}}$. 
These evolutionary laws show that the number and luminosity density of HLAGN increased from higher redshifts ($\mathrm{z\sim6}$) up to a maximum in the redshift range $ \mathrm{1 < z < 2.5}$, followed by a decline towards local values. 
By scaling the 1.4 GHz AGN luminosity to kinetic luminosity using the standard conversion, we estimate the kinetic luminosity density as a function of redshift. We compare our result to the semi-analytic models of radio mode feedback finding that this feedback could have played an important role in the context of AGN-host coevolution in HLAGN which show evidence of AGN-related radio emission ($\mathrm{f_{AGN}>0}$).
}		

\maketitle
\makeatother

\section{Introduction\label{sec:intro}}

The coevolution of the host galaxies and the active galactic nuclei (AGN) within them has been widely accepted after the discovery of scaling relations between supermassive black hole (SMBH) mass and host galaxy properties (e.g. \citealt{magorrian98}, \citealt{fm00}). To better understand this coevolution we need to  examine how AGN and their host galaxies change through cosmic time.

AGN are powerful extragalactic sources visible across the entire electromagnetic spectrum, from gamma-rays and X-rays to the radio. In order to study their activity and how it changes through cosmic time a multiwavelength approach is needed. 
Observations show that AGN activity occurs in at least two different flavors, resulting in two intrinsically different AGN populations (\citealt{ss73}, \citealt{ny94}, \citeyear{ny95}). This dichotomy most likely occurs due to the difference in the efficiency of the mechanism that governs the accretion onto the central SMBH (e.g. see review by \citealt{hb14}). 
In the case of inefficient accretion fueled by the inflow of the hot gas at low accretion rates (below $\mathrm{1\%}$ of Eddington), the main energetic output is the kinetic energy transported by powerful radio jets, which can outshine the radio emission from the host galaxy. They produce a relatively small amount of radiation across the rest of the electromagnetic spectrum and are hence called low-excitation radio galaxies (LERGs) or jet-mode AGN (e.g. \citealt{bh12}, \citealt{best14}, \citealt{pracy16}). The cold gas accretion at high rates (above $\mathrm{1\%}$ of Eddington) onto the central SMBH is radiatively efficient and can produce signatures detectable as optical emission lines. Objects containing this type of signature are referred to as high-excitation radio galaxies (HERGs) or radiative-mode AGN, some of which are found to also have strong radio jets.
AGN can also be classified based on their radio power as radio-quiet (RQ) and radio-loud (RL) AGN. If this classification is based on the ratio of radio and infrared luminosities (as in, e.g., \citealt{padovani15}), most HERGs/radiative-mode AGN are found to be radio-quiet, while LERGs/jet-mode AGN are radio-loud due to the presence of powerful radio jets (\citealt{smolcic16}).

The difference between the two populations is reflected also in the shape of the radio luminosity functions constructed from local samples of radio galaxies (\citealt{filho06}, \citealt{bh12}, \citealt{gendre13}, \citealt{best14}, \citealt{pracy16}). The jet-mode AGN and LERGs dominate at lower radio luminosities, while radiative-mode AGN and HERGs become dominant at radio luminosities higher than $\mathrm{L_{1.4\ GHz} \sim 10^{26}\ WHz^{-1}}$. However, there is substantial disagreement at $\mathrm{L_{1.4\ GHz}<10^{25}\ WHz^{-1}}$ between luminosity functions derived from various surveys of the nearby Universe, such as Faint Images of the Radio Sky at Twenty-cm (FIRST), NRAO VLA Sky Survey (NVSS), Sloan Digital Sky Survey (SDSS), Palomar Sky Survey (\citealt{becker94}, \citealt{condon98}, \citealt{york00}, \citealt{palomar97}, respectively). 
While \citeauthor{filho06} (\citeyear{filho06}; based on the Palomar Sky Survey) and \citeauthor{pracy16} (\citeyear{pracy16}; based on the Large Area Radio Galaxy Evolution Spectroscopic Survey - LARGESS) find substantial volume densities at luminosities $\mathrm{L_{1.4\ GHz}\lesssim 10^{25}\ WHz^{-1}}$, \citeauthor{bh12} (\citeyear{bh12}; based on the SDSS survey) find up to about an order of magnitude lower volume densities at these luminosities (see Fig. 7 in \citealt{pracy16}). As argued by \citeauthor{pracy16}, this disagreement may be due to the different approaches used to eliminate star-forming galaxies from the HERG sample. \citet{bh12} chose a conservative approach, removing all radio galaxies where the radio emission was suspected to arise from star formation, and this may have led to an underestimation of the volume densities of their HERG population (particularly at the low-luminosity end). Vice versa, the contribution from star-forming processes may bias the low luminosity end of the luminosity functions as derived by \citet{filho06} and \citet{pracy16}.

Due to the difference in the supply of cold gas needed for triggering star formation, the two different types of AGN are hosted by galaxies with different stellar populations (e.g. \citealt{smolcic09}). For radio galaxies out to redshift $\mathrm{z\sim 1}$ it has been found that HERGs have bluer colors associated with ongoing star formation, while LERGs are mostly red quiescent galaxies (e.g. \citealt{smolcic09}, \citealt{bh12}, \citealt{ching17}). While this might not hold for the highest luminosity radio galaxies (see, e.g., \citealt{dunlop03}), a study by \citet{bonzini13}, which, because of the small covered area, does not sample the brightest part of the radio luminosity function, found that the RQ and RL AGN with redshifts 0.1 < z < 4 are hosted mainly in late- and early-type galaxies, respectively. To constrain the evolution of an AGN component, one first needs to infer how much of the detected radio emission arises from AGN activity and how much from star-forming processes. While the radio emission in the jet-mode AGN is expected to be dominated by radio jets, the host properties of radiative-mode AGN indicate that a significant fraction of their radio luminosity arises from ongoing star formation. Most radio sources detected in surveys such as FIRST, SDSS, Cosmic Evolution Survey (COSMOS) and the Extended Chandra Deep Field South (E-CDFS) are unresolved and morphological decomposition of radio emission into AGN and star-forming components cannot be performed. In order to study the evolution of these components separately, another approach is needed, as taken here. 

Studies of the evolution of spectroscopically selected AGN show different evolutionary trends; while LERGs show little or no evolution, HERGs seem to be a strongly evolving population (e.g. \citealt{bh12}, \citealt{pracy16}). \citet{best14} studied the evolution of a sample of over 200 sources spectroscopically classified as radiative- or jet-mode AGN from a combination of eight radio surveys. They found an increase in the space density of the radiative-mode AGN from the local Universe out to redshift $z\mathrm{\sim 0.75}$. For the lower luminosity ($\mathrm{L_{1.4\ GHz}\leq 10^{25} \mathrm{WHz^{-1}}}$) jet-mode AGN they find no evidence of evolution, while their higher radio luminosity counterparts showed a significant increase in space density up to $\mathrm{z\sim 1}$. Similar results were found in the study of the radio sample of 680 objects from the E-CDFS (\citealt{padovani15}), classified as RL and RQ AGN based on the ratio between 24 $\mathrm{\mu}$m and 1.4 GHz flux densities. Their RL AGN show an increase in number density out to a peak at $\mathrm{z\sim0.5}$, followed by a decline towards the higher redshifts. The population of RQ AGN shows a more rapid evolution, with a hint of a slower evolution at redshifts $\mathrm{z>1.3}$. 
As demonstrated by \citet{rigby15}, there exists a luminosity-dependence of the redshift evolution of the radio luminosity function. Using a combination of several radio datasets they have selected the AGN as sources which show radio excess relative to that expected from star formation, using the so-called q parameter, defined as the ratio of the mid-infrared ($\mathrm{24\ \mu m}$) to radio ($\mathrm{1.4\ GHz}$) flux density. Their results indicate that the peak of the space density appears at higher redshifts for the most powerful sources ($\mathrm{L_{1.4\ GHz} \geq 10^{25.6}\ WHz^{-1}}$), while for their lower power counterparts ($\mathrm{L_{1.4\ GHz} < 10^{25.6}\ WHz^{-1}}$) the space density appears to be constant with redshift. Their interpretation of the results is that the higher power sources represent the cold-mode sources which were the more dominant population in the early Universe, while the plateau at lower powers corresponds to the change in dominance between the cold- and hot-mode sources towards the local Universe.

The feedback mechanism that governs the coevolution between an AGN and the host galaxy is thought to occur in two flavors, namely radio and quasar mode feedback. The radio mode feedback was first introduced to solve the cooling flow problem arising in the earliest semi-analytical models and numerical simulations of formation of structure (e.g. \citealt{bower06}, \citealt{croton06}, \citeyear{croton16}). It operates in massive galaxies where powerful radio jets inject energy in kinetic form into the surrounding environment, effectively heating the gas and preventing star formation. On the other hand, the quasar mode feedback is associated with periods of rapid black hole growth due to the high accretion rates triggered by galaxy mergers or disk instabilities. These episodes of rapid accretion produce quasar winds which can affect the surrounding gas blowing it out of the galaxy and hence also terminating the formation of new stars (e.g. \citealt{croton16}).

The kinetic energy injected into the environment can be estimated using one of many scaling relations between monochromatic radio luminosity and kinetic luminosity (e.g. \citealt{willott99}, \citealt{mh07}, \citealt{cavagnolo10}, \citealt{gs16}). 
However, it is important to note that different scaling relations can result in estimates of kinetic luminosity which differ by several orders of magnitude (see Appendix A in \citealt{smolcic17c} for details). Keeping this in mind, the estimates of the kinetic luminosity density can be further compared to models of galaxy evolution (e.g. \citealt{bower06}, \citealt{croton06}, \citeyear{croton16}, \citealt{mh08}). 
A recent study of the LERG and HERG evolutions by \citet{pracy16} out to $\mathrm{z\sim 0.75}$ found little or no evolution of the kinetic luminosity density for the most powerful LERGs, which is consistent with the result for the radio-mode AGN from the semi-analytical model of galaxy evolution by \citet{croton06}. 
The local value of the kinetic luminosity density of HERGs is an order of magnitude lower than that of LERGs, but becomes comparable at $\mathrm{ z\sim 0.75}$, showing that the radio mode feedback in HERGs influenced their evolution more at higher redshift than it does in the local Universe. 

{In this work we study the cosmic evolution of a population of AGN in the COSMOS field, detected in the VLA-COSMOS 3 GHz Large Project out to $\mathrm{z\sim6}$. Selection of these AGN is based on a combination of X-ray and mid-infrared (MIR) criteria and broad-band spectral energy distribution fitting. The applied selection is aimed to trace the analogs of HERGs, as these criteria are sensitive to the excess of emission likely to arise due to radiatively efficient accretion onto SMBH. We call these sources moderate-to-high radiative luminosity AGN (HLAGN; see Sec. \ref{sec:hlagn}). 
We develop a statistical method for decomposing the total radio luminosity into the star forming and AGN components to trace the evolution of radio emission associated only with the AGN.

The paper is structured as follows. In Sec. \ref{sec:data} we describe the data set used in the analysis. In Sec. \ref{sec:sf} we investigate star formation in the HLAGN population and describe the procedure adopted to decompose the total 1.4 GHz luminosity into contributions arising from an AGN and from processes related to star formation. AGN luminosity functions and their evolution are derived and described in Sec. \ref{sec:luminosity_function} and Sec. \ref{sec:lf_evolution}, respectively. In Sec. \ref{sec:discussion} we discuss our results in the context of studies from the literature in order to understand what the origin of AGN-related radio emission in HLAGN is and how it may influence the properties of their host galaxies. Finally, in Sec. \ref{sec:sac} we summarize the conclusions of our work.

Throughout this paper we assume $\mathrm{\Lambda CDM}$ cosmology with $\mathrm{H_0 = 70\ km s^{-1} Mpc^{-1}}$, $\mathrm{\Omega_m = 0.3}$ and $\mathrm{\Omega_{\Lambda} = 0.7}$.
We assume that the synchrotron radio emission follows a simple power law spectrum of the form $\mathrm{S_\nu \propto \nu^{\alpha}}$, where $\mathrm{S_\nu}$ is the flux density at frequency $\mathrm{\nu}$ and $\mathrm{\alpha}$ is the spectral index.

\section{Data and sample} \label{sec:data}

\subsection{The VLA-COSMOS 3 GHz Large Project} \label{sec:cosmos}

To investigate the evolution of the radio-AGN luminosity function over cosmic time, a large sample of sources with available multiwavelength data and high-quality redshifts is needed. The VLA-COSMOS 3 GHz Large Project provides radio data from 384 hours of observations with the Karl G. Jansky Very Large Array (VLA). The 3 GHz (10 cm) observations were carried out over $\mathrm{2.6\ deg^{2}}$ centered in the COSMOS field, reaching a median $\mathrm{1\sigma}$ sensitivity of $\mathrm{2.3\ \mu Jy / beam}$ over the inner $\mathrm{2\ deg^{2}}$ at an angular resolution of  $\mathrm{0.75"}$. The total number of radio detected sources with signal-to-noise ratio (S/N) $\mathrm{\geq 5}$ across the entire $\mathrm{2\ deg^{2}}$ COSMOS field is 10,830. All details of the observations and the extraction of radio sources can be found in \citet{smolcic17a}. 

\citet{smolcic17b} cross-matched the VLA-COSMOS 3 GHz Large Project radio data with the multiwavelength COSMOS2015 catalog (\citealt{laigle16}), identifying a sample of 7,729 radio sources with optical/near IR counterparts. This is $\mathrm{89\%}$ of all radio sources in the effective area considered ($\mathrm{1.77\ deg^2}$). For $\mathrm{\sim 31 \%}$ ($\mathrm{\sim69\%}$) of these sources, reliable spectroscopic (photometric) redshifts are available. $\mathrm{33\%}$ of radio sources have also been detected in the 1.4 GHz VLA-COSMOS Large Survey (\citealt{schinnerer07}, \citeyear{schinnerer10}). For these sources the spectral index is calculated from the observed slope between 1.4 and 3 GHz. Otherwise, the spectral index is set to be -0.7 which is the expected value for non-thermal synchrotron radiation (e.g. \citealt{condon92}) and corresponds to the average spectral index found for 3 GHz sources in this survey (\citealt{smolcic17a}).

\subsection{Identification of moderate-to-high radiative luminosity AGN (HLAGN)} \label{sec:hlagn}
The radio sources detected in the VLA-COSMOS 3 GHz Large Project have been separated into the following classes by \citet{smolcic17b}: low-to-moderate radiative luminosity AGN (MLAGN), moderate-to-high radiative luminosity AGN (HLAGN) and star-forming galaxies (SFGs). We briefly summarize the separation method here. 
Firstly, HLAGN have been identified using a combination of X-ray\footnote{The X-ray data used to classify HLAGN are taken from the COSMOS Legacy/Chandra COSMOS surveys (see \citealt{civano16}, \citealt{marchesi16}).} ($\mathrm{L_X > 10^{42}\ ergs^{-1}}$) and MIR\footnote{The MIR data used to select HLAGN are taken from the SPLASH project (IRAC/Spitzer space telescope; see \citealt{steinhardt14}).} (color-color diagram; see \citealt{donley12}) criteria and via template fitting to the optical-to-millimeter spectral energy distributions (SED; see \citealt{delvecchio17}).
The union of these three criteria yielded a sample of 1,604 sources classified as HLAGN\footnote{A justification for the use of the HLAGN nomenclature is provided in \citet{smolcic17b}. They find that, despite the overlap between two classes, HLAGN sources classified as such on the basis of X-ray and MIR criteria, as well as SED fitting, tend to have higher ($\mathrm{\sim1\ dex}$) radiative luminosities than MLAGN. This trend is present at all redshifts they studied ($\mathrm{0.01\leq z \leq 5.70}$), confirming that this naming convention is justified in a statistical sense (see also \citealt{delvecchio17}).}. These objects are the main focus of this paper. The other radio sources were classified either as SFGs or MLAGN, based on optical colors, Herschel imaging and radio emission.  

For every radio source, infrared (IR) luminosities arising from star-forming processes were obtained from the best-fit galaxy SED, after correcting for a possible AGN-related emission (for more details see \citealt{delvecchio17}). Star formation rates (SFRs) were calculated from these luminosities, by assuming the \citet{kennicutt98} scaling relation and a \citet{chabrier} initial mass function. The data products as described by \citet{smolcic17a} and \citet{delvecchio17} are available at the COSMOS IPAC/IRSA archive\footnote{http://irsa.ipac.caltech.edu/data/COSMOS/tables/vla/}.

\section{Star formation properties of HLAGN} \label{sec:sf}

In this section we examine the star formation contribution to the radio emission of HLAGN. We present a statistical decomposition method developed to separate the total radio luminosity into contributions originating from AGN activity and star formation.

\subsection{Star formation in HLAGN host galaxies} \label{sec:sf_hlagn}

Non-thermal radio emission in extragalactic surveys arises either from AGN or star-forming processes (e.g. \citealt{condon92}, \citealt{miller93}). \citet{delvecchio17} analyzed the properties of MLAGN and HLAGN (see Sec. \ref{sec:hlagn}). They find that MLAGN at redshifts $\mathrm{z\leq1}$ show SFRs and stellar masses consistent with those of passive galaxies with little or no ongoing star formation. For the HLAGN population they find that the majority lies on the main sequence for star formation (\citealt{whitaker12}). They also argue that about $\mathrm{70\%}$ of HLAGN exhibit no `significant' ($\mathrm{>3\sigma}$) radio-excess relative to their IR-based SFRs, suggesting that a large fraction of radio emission in HLAGN is of star formation origin.

\begin{figure*}[!t]
  \centering
   \includegraphics[ width=\linewidth]{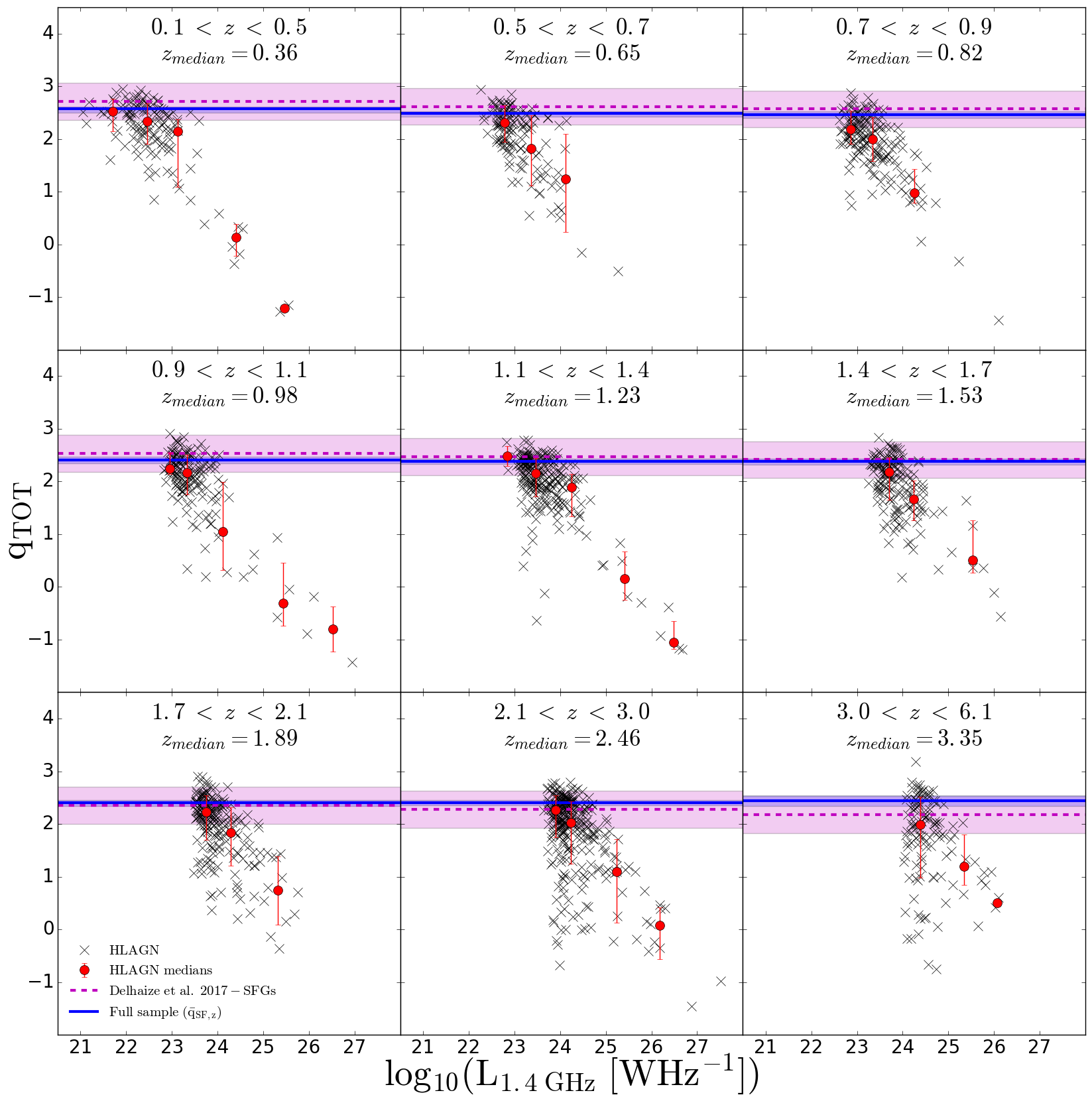}
  \caption{The infrared-to-1.4 GHz radio luminosity ratio ($\mathrm{q_{TOT}}$) vs. $\mathrm{L_{1.4\ GHz}}$ for HLAGN (black crosses). Radio luminosity binned median values of the ratio are shown with red circles with error bars showing 16th and 84th percentiles of the distribution. The dashed magenta line corresponds to the IRRC derived by \citet{delhaize17} at the median redshift of the underlying HLAGN sample, along with its $\mathrm{1\sigma}$ spread of $\mathrm{\pm0.35\ dex}$. Blue solid lines and area show the median of the star-forming peak $\mathrm{\bar{q}_{SF,z}}$ and $\mathrm{1\sigma_{q_{SF,z}}}$ uncertainties calibrated on the full radio detected sample of the VLA-COSMOS 3 GHz Large Project sources with the COSMOS2015 counterparts.}
  \label{fig:plot_qz}
\end{figure*}

Both the infrared and radio emissions can be used to trace the recent star formation in SFGs, which leads to a tight correlation between them, the so-called infrared-radio correlation (IRRC; e.g. \citealt{yun01}, \citealt{bell03}, \citealt{magnelli15}).
\citet{delhaize17} examined the cosmic evolution of the IRRC in the COSMOS field using a sample of 9,575 SFGs. This sample was jointly selected in the far-infrared (based on Herschel fluxes) and in the radio (objects in the VLA-COSMOS 3 GHz sample) after removing HLAGN and quiescent MLAGN (as described by \citealt{smolcic17b}). They found a redshift-dependent infrared-to-1.4 GHz radio luminosity ratio, $\mathrm{q_{TIR}(z)}$, with the $\mathrm{1\sigma}$ spread of $\mathrm{0.35\ dex}$. The spread around the IRRC shows the part of the IRRC parameter space in which we expect most of the radio emission to be produced in star-forming processes (e.g. \citealt{yun01}, \citealt{bell03}).

In Fig. \ref{fig:plot_qz} we show the redshift evolution of the logarithm of infrared-to-total 1.4 GHz luminosity ($\mathrm{q_{TOT}}$) vs. the total 1.4 GHz luminosity ($\mathrm{L_{1.4\ GHz, TOT}}$) of the HLAGN sample, overlaid with the \citeauthor{delhaize17} $\mathrm{q_{TIR}(z)}$ result. A significant number of HLAGN are found within and above the $\mathrm{1\sigma}$ envelope around the IRRC at all redshifts, which is qualitatively consistent with the results by \citet{delvecchio17} and \citet{smolcic17b}. This implies that star-forming processes contribute considerably to the radio emission of these HLAGN, while the excess of the radio emission in HLAGN below the IRRC envelope is mainly powered by the AGN activity. To quantify the extent to which star-forming processes and AGN activity contribute to the total 1.4 GHz luminosity, we perform a decomposition analysis as described in the next section.

\subsection{Decomposition of star formation and AGN contributions to the total radio luminosity} \label{sec:decomposition}

Most of the previous studies of luminosity functions have been conducted using surveys at 1.4 GHz (e.g. \citealt{bh12}, \citealt{pracy16}). To compare to and complement those studies, we here derive 1.4 GHz radio luminosities. From the observed 3 GHz flux densities ($\mathrm{S_{3\ \mathrm{GHz},TOT}}$; $\mathrm{WHz^{-1}m^{-2}}$) we calculate the rest frame radio luminosity at 1.4 GHz ($\mathrm{L_{1.4\ \mathrm{GHz,TOT}}}$; $\mathrm{WHz^{-1}}$) using:

\begin{equation}\label{eq:14ghz_luminosity}
$$\mathrm{L_{1.4\ GHz, TOT} = \dfrac{4\pi D_{L}^{2}(z)}{(1+z)^{1+\alpha}}\bigg(\dfrac{1.4\ \mathrm{GHz}}{3\ \mathrm{GHz}}\bigg)^{\alpha}S_{3\ \mathrm{GHz}, TOT}},$$
\end{equation}
where $\mathrm{D_{L}}$ and $\mathrm{z}$ are the luminosity distance (in unit $\mathrm{m}$) and redshift of the source, respectively. 
The index ‘TOT’ refers to the total 1.4 GHz radio luminosity and the 3 GHz flux density prior to decomposition. As previously discussed, we expect the radio emission in HLAGN to originate from both star-forming processes in the host galaxy and AGN activity. 
The total 1.4 GHz radio luminosity ($\mathrm{L_{1.4\ GHz, TOT}}$) is hence the sum of star forming ($\mathrm{L_{1.4\ GHz, SF}}$) and AGN ($\mathrm{L_{1.4\ GHz, AGN}}$) contributions. To quantify these contributions, we perform the decomposition analysis described here. 

\begin{figure}[!h]
\includegraphics[width=\linewidth]{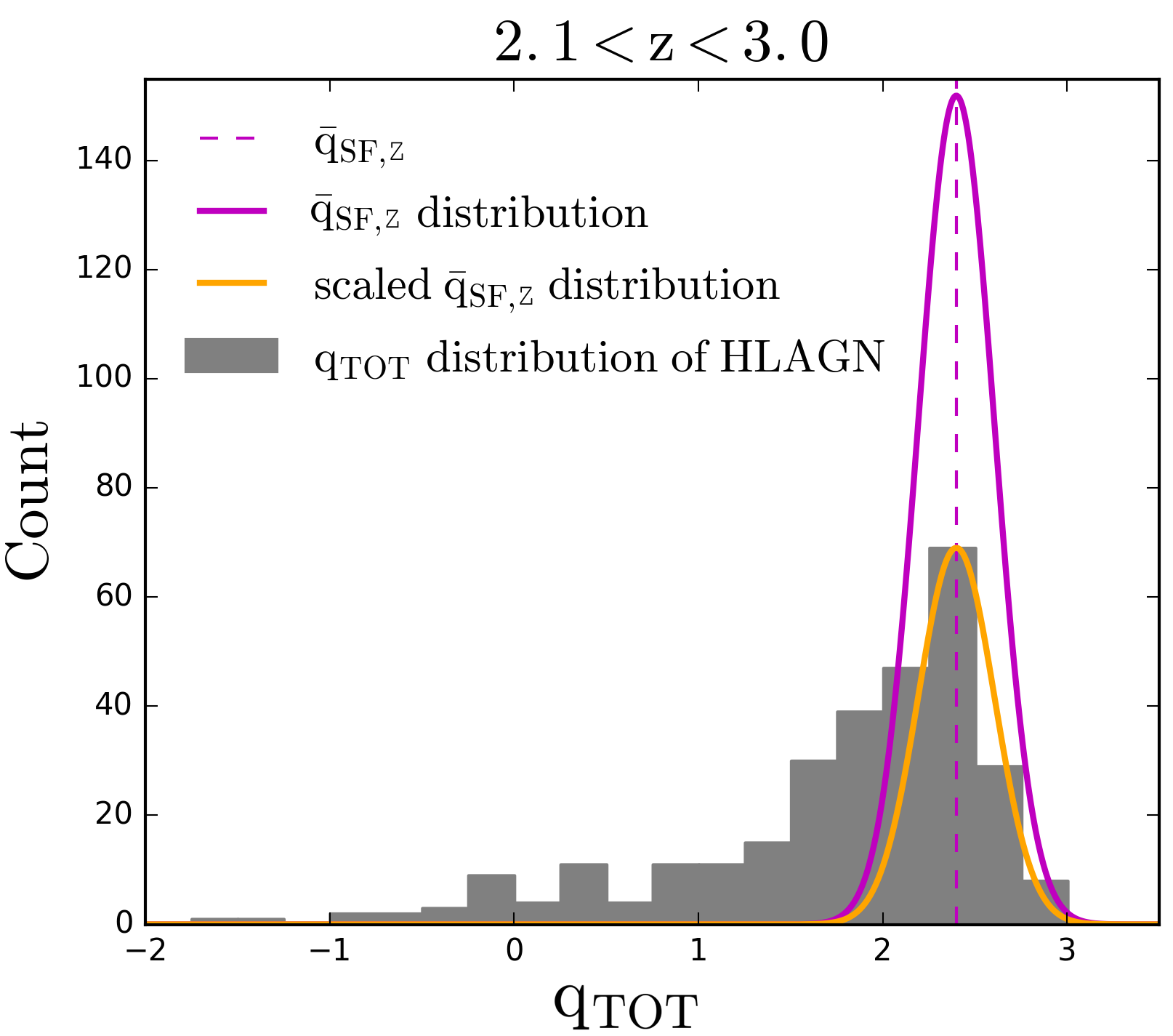}
\caption{An example of one iteration of the $\mathrm{q_{TOT}}$ distribution for the HLAGN (gray histogram) overlaid with the Gaussian functional forms showing the contribution of star-forming processes in the redshift bin $\mathrm{2.1<z<3.0}$. Gaussian functional forms show the contribution from the star formation as derived from the full sample and after scaling it to the normalization of the HLAGN distribution at the $\mathrm{\bar{q}_{SF,z}}$ (shown with magenta and orange lines, respectively).}
\label{fig:distributions}
\end{figure}

We split our HLAGN sample into nine redshift bins, each with roughly equal number of sources, in the redshift range $\mathrm{0.1<z<6.1}$. We calculate the infrared-to-total 1.4 GHz luminosity ratio:
\begin{equation}\label{eq:qtot}
$$\mathrm{q_{TOT}\propto log{\frac{L_{IR, SF}}{L_{1.4\ GHz, TOT}}}},$$
\end{equation}
where $\mathrm{L_{IR,SF}}$ is the infrared luminosity only due to star-forming processes (see Sec. \ref{sec:hlagn}). Uncertainties on $\mathrm{q_{TOT}}$, $\mathrm{\sigma_{q_{TOT}}}$, are estimated by propagating the uncertainties on the luminosities. Assuming for each source a normal distribution of $\mathrm{q_{TOT}}$ with root mean square scatter $\mathrm{\sigma_{q_{TOT}}}$, and drawing $\mathrm{q_{TOT}}$ from this distribution, in Fig. \ref{fig:distributions} we show one representation of the $\mathrm{q_{TOT}}$ distribution for our HLAGN sample ($\mathrm{2.1<z<3.0}$; gray histogram in Fig. \ref{fig:distributions}).

We next compute the median $\mathrm{q_{SF}}$ calibrated on the full radio-detected sample in each redshift bin ($\mathrm{\bar{q}_{SF,z_i}}$). Here, $\mathrm{q_{SF}}$ is the infrared-to-1.4 GHz radio luminosity expected to arise only from star-forming processes, with the corresponding uncertainties $\mathrm{\sigma_{SF}}$. Appendix \ref{sec:comp_c15} describes in detail how $\mathrm{\bar{q}_{SF,z}}$ was calculated. 

We assume that for HLAGN, the star formation contribution to the radio emission follows the same Gaussian functional form expected for the star-forming galaxy sample of radio detected sources; i.e. that it is peaked at $\mathrm{\bar{q}_{SF, z}}$ with the corresponding dispersion $\mathrm{\sigma_{q_{SF, z}}}$. We set the normalization of the Gaussian to the value of the HLAGN $\mathrm{q_{TOT}}$ distribution at the position of the $\mathrm{\bar{q}_{SF,z}}$ (see purple and orange lines in Fig. \ref{fig:distributions}).

\begin{figure}[!h]
\includegraphics[width=\linewidth]{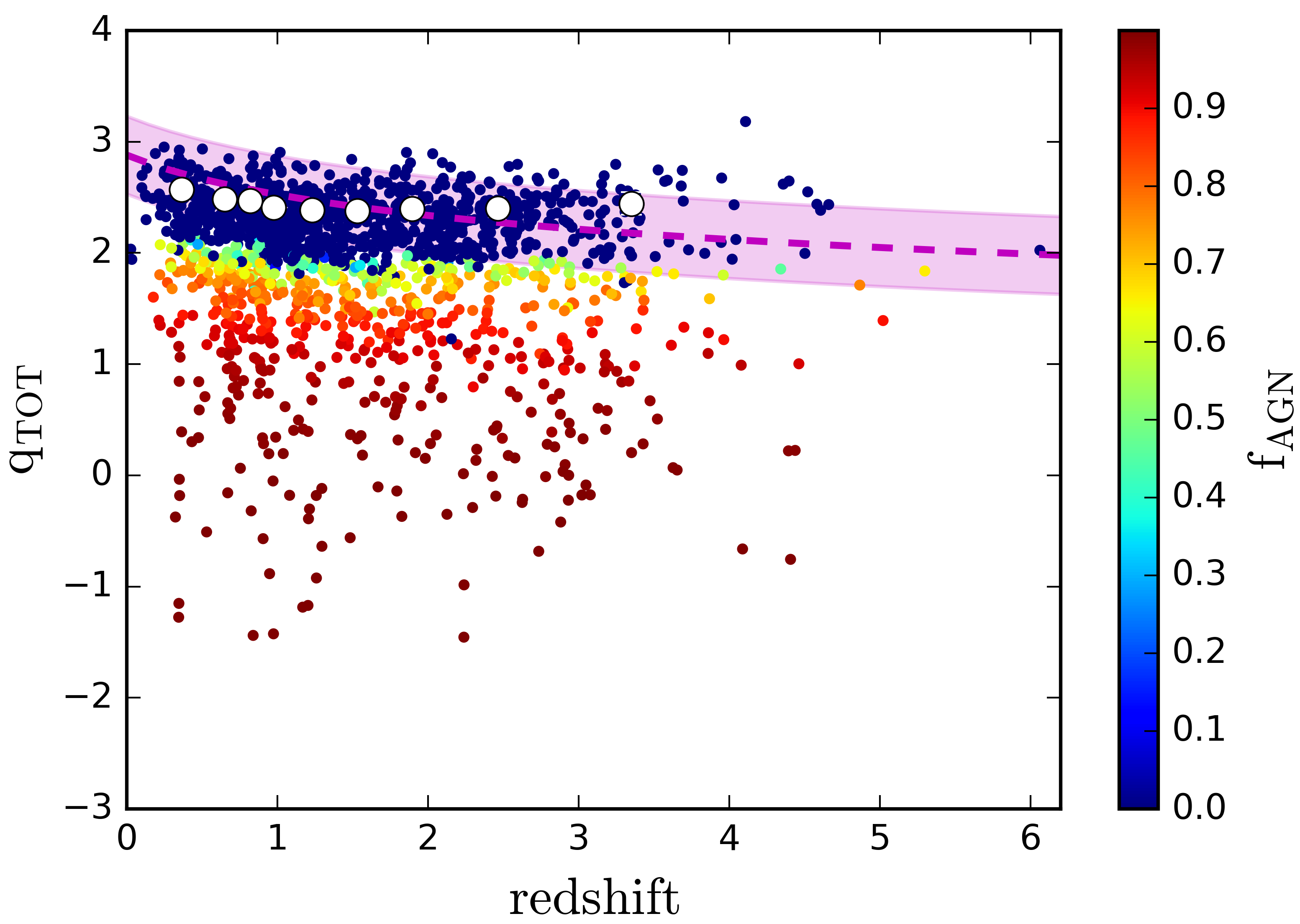}
\caption{The infrared-to-1.4 GHz radio luminosity ratio ($\mathrm{q_{TOT}}$) vs. redshift for the HLAGN. The data are color-coded by their median values of AGN fractions. The magenta dashed line is the redshift dependent analytic form calibrated by \citet{delhaize17} on a sample of SFGs, with $\mathrm{\pm 1\sigma}$ spread (magenta area). White dots show the peaks ($\mathrm{\bar{q}_{SF,z}}$) of the underlying $\mathrm{q_{TOT}}$ distribution calibrated on the star-forming galaxies in the full radio detected sample.}
  \label{fig:q_vs_z_hlagn}
\end{figure}
 
Then we find the difference in number count between the observed $\mathrm{q_{TOT}}$ distribution of HLAGN (gray filled histogram in Fig. \ref{fig:distributions}) and the normalized $\mathrm{\bar{q}_{SF}}$ distribution (orange line in Fig. \ref{fig:distributions}) in bins of $\mathrm{q_{TOT}}$\footnote{To minimize the effect binning might have on our results, we use Freedman-Diaconis rule (\citealt{fd81a}) to choose $\mathrm{q_{TOT}}$ bin widths: $$\mathrm{Bin\ size = 2\ IQR(x)\ n^{-1/3}},$$ where $\mathrm{IQR(x)}$ is the interquartile range of the data within the bin and $\mathrm{n}$ is the number of sources within that bin.}. This gives an estimate of the number of sources ($\mathrm{N_{AGN}}$) expected to have AGN emission contributing to the total radio emission. We randomly select $\mathrm{N_{AGN}}$ sources within each $\mathrm{q_{TOT}}$ bin and calculate their AGN fractions at 1.4 GHz, $\mathrm{f_{AGN}}$, defined as:
\begin{equation} \label{eq:f_agn}
$$\mathrm{f_{AGN} = 1-10^{(q_{TOT}-\bar{q}_{SF,z})}}.$$
\end{equation}
If a source within a particular $\mathrm{q_{TOT}}$ bin is not selected in this process, it is given a $\mathrm{f_{AGN}=0}$ value in this particular iteration.
To account for the uncertainties, we repeat this entire procedure $\mathrm{1,000}$ times using Monte Carlo simulations, constructing a new $\mathrm{q_{TOT}}$ distribution each time. 
The result is a distribution of $\mathrm{f_{AGN}}$ values for each source, which accounts for the distribution of $\mathrm{q_{SF, z}}$ and the uncertainty on $\mathrm{q_{TOT}}$. 

Fig. \ref{fig:q_vs_z_hlagn} shows the distribution of $\mathrm{q_{TOT}}$ as a function of redshift for our HLAGN sample, color-coded by the derived AGN fractions. AGN fractions range from 0 to 1, where the two extreme values correspond to situations where all radio emission is due to the star-forming processes ($\mathrm{f_{AGN} = 0}$) or AGN activity ($\mathrm{f_{AGN} = 1}$). Given the definition of $\mathrm{f_{AGN}}$ in Eq. \ref{eq:f_agn}, AGN fractions can assume unphysical (negative) values if the source has randomly extracted $\mathrm{q_{TOT}}$ value greater than the average $\mathrm{q_{SF,z}}$ value calibrated on the full radio sample. In these cases we set $\mathrm{f_{AGN}=0}$ which is equivalent to assuming that radio emission arises entirely from star formation.
The distribution of the median AGN fractions is shown in Fig. \ref{fig:hlagn_distributions}.

\begin{figure}[!h]
  \includegraphics[width=\linewidth]{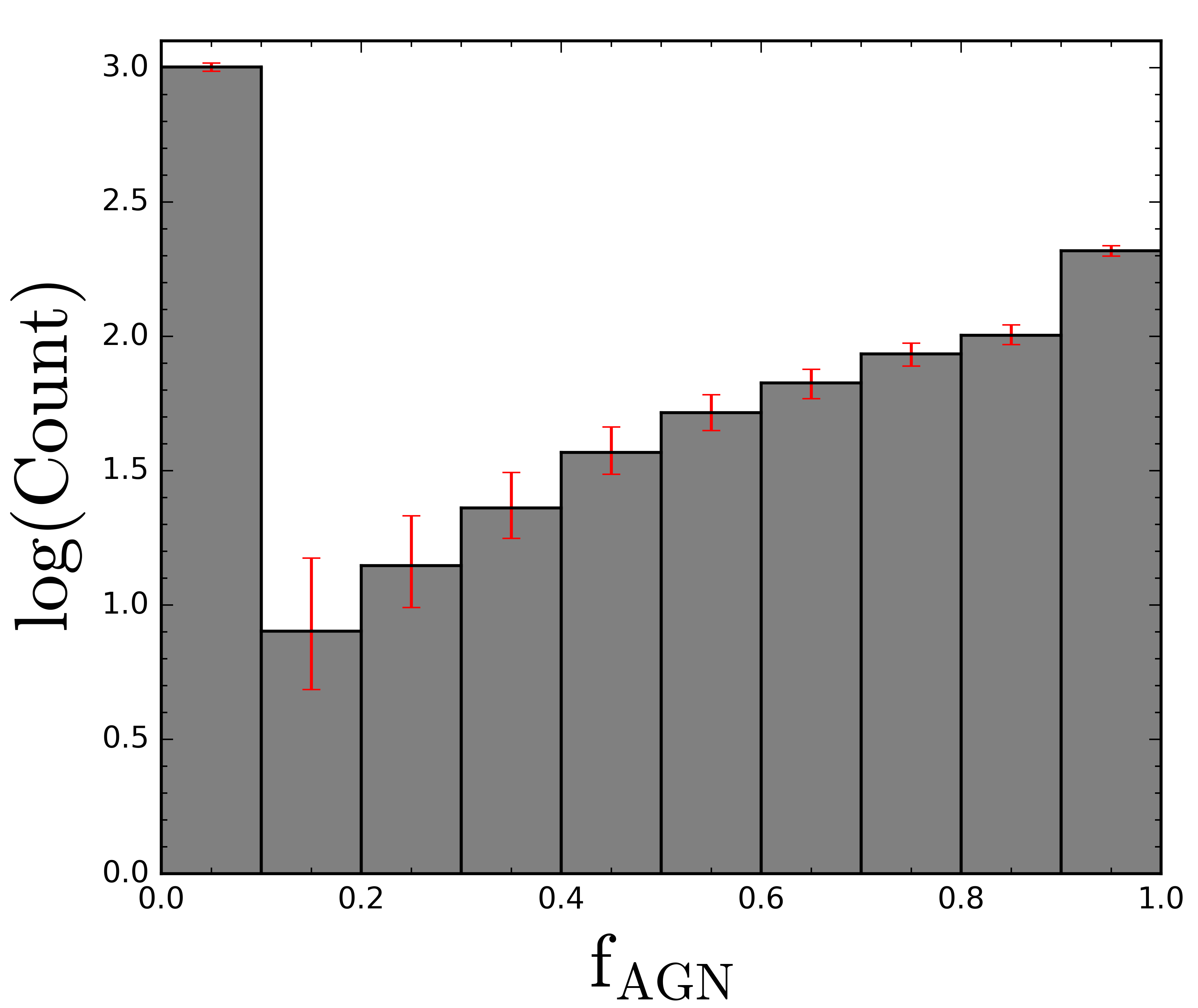}
  \caption{Histogram of median AGN fractions across all redshifts, computed using $\mathrm{Eq. \ref{eq:f_agn}}$ (see text for details). Error bars show the 16th and 84th percentile values.}
  \label{fig:hlagn_distributions}
\end{figure}

We find that the majority of HLAGN, $\mathrm{(68.0 \pm 1.5) \%}$, have the total radio emission dominated by star-forming processes ($\mathrm{0\leq f_{AGN}\leq0.5}$), which is consistent with results by \citet{smolcic17b} and \citet{delvecchio17}. On the other hand, a subsample of HLAGN is dominated by the AGN-related radio emission. We find that $\mathrm{(9.9\pm 0.9)\%}$ of HLAGN have 1.4 GHz AGN fractions in the range $\mathrm{0.5 < f_{AGN} \leq 0.75}$, while $\mathrm{(22.1 \pm 0.9)\%}$ have these fractions of $\mathrm{0.75 < f_{AGN} \leq 1}$.  

We scale the total 1.4 GHz radio luminosity and the observed 3 GHz flux density down to the AGN-related contributions:
\begin{equation}\label{eq:agn_luminosity}
$$\mathrm{L_{1.4\ GHz, AGN}=L_{1.4\ GHz, TOT}\ f_{AGN},}$$
\end{equation}
\begin{equation}\label{eq:agn_flux}
$$\mathrm{S_{3\ GHz, AGN}=S_{3\ GHz, TOT}\ f_{AGN}.}$$
\end{equation}

For the $\mathrm{(37.5\pm2.2)\%}$ of HLAGN that show evidence for AGN-related radio emission ($\mathrm{f_{AGN}>0}$), we estimate the 1.4 GHz AGN luminosities via the decomposition method. To test whether AGN fractions derived for this subsample evolve with redshift, we derive the median and $\mathrm{1 \sigma}$ values per redshift bin, but find no significant evidence of evolution. The median of the AGN fraction of the HLAGN sample with $\mathrm{f_{AGN}\neq0}$ is $\mathrm{0.83^{+0.15}_{-0.23}}$. 

Only the radio luminosity component due to AGN ($\mathrm{L_{1.4\ GHz, AGN}}$) is further used to calculate the radio AGN luminosity functions. This effectively allows us to exclude `contamination' of the AGN luminosity functions by star formation-related emission and thus derive the genuine radio AGN luminosity function.

\section{Radio luminosity functions of HLAGN} \label{sec:luminosity_function}

In this section we describe the derivation of radio AGN luminosity functions. 
To ensure we have a complete sample of sources with fully observable AGN radio emission by our survey, we impose a cut in the AGN flux densities $\mathrm{S_{3\ GHz, AGN}\geq S_{lim, 3\ GHz}}$, where $\mathrm{S_{lim, 3\ GHz}}$ is the detection limit of the survey ($\mathrm{S_{lim, 3\ GHz}=5\sigma=11.5 \mu Jy}$). Then we construct the AGN luminosity functions using the AGN luminosities obtained with the decomposition technique for such defined sample of $575$ HLAGN.
To show how the decomposition affects the shape of the luminosity functions, we also calculate the total luminosity functions for the full HLAGN sample \textbf{($\mathrm{0 \leq f_{AGN}\leq 1.0}$)}, using the total 1.4 GHz luminosity which includes emission of both AGN and star-formation origin. Finally, we compare our results with the literature.

\subsection{Derivation} \label{sec:lf_derivation}

To constrain the redshift evolution of the radio AGN luminosity function we calculate the comoving space density of radio sources per logarithm of luminosity in nine redshift bins (same binning as in Sec. \ref{sec:decomposition}). To find the maximum observable volume in which each of the galaxies in our sample could be observed we use the $\mathrm{V_{max}}$ method (\citealt{schmidt68}):
\begin{equation}\label{eq:vmax}
$$\mathrm{V_{max, i}= \int_{z_{min}}^{z_{max}}\mathcal{C}(S_{3\ GHz, AGN})\dfrac{dV}{dz} dz},$$
\end{equation}
where $\mathrm{V_{max,i}}$ is the maximum observable volume of the i-th galaxy in the sample, $\mathrm{z_{min}}$ is the minimum redshift of the redshift bin under study, and $\mathrm{z_{max}}$ is the maximum redshift where the source can still be part of the sample given its radio luminosity. 
To take into account the area covered by the radio data and the variation of completeness of the radio catalog as a function of the observed flux, we define the completeness factor ($\mathcal{C}$) as:

\begin{equation}\label{eq:correction}
$$\mathrm{\mathcal{C}(S_{3\ GHz, AGN}) = \dfrac{A_{obs}}{A_{sky}} \times \mathcal{C}_{comp}(S_{3\ GHz, AGN})},$$
\end{equation}
where $\mathrm{A_{obs} = 1.77\ deg^{2}}$ is the effective unmasked area of the COSMOS optical-NIR survey, $\mathrm{A_{sky}}$ is the area of the sky sphere, and $\mathrm{\mathcal{C}_{comp}(S_{3\ GHz, AGN})}$ is the completeness of the radio catalog at the $3$ GHz AGN flux density $\mathrm{S_{3\ GHz, AGN}}$ (for the completeness of the survey, see Fig. 16 and Table 2 in \citealt{smolcic17a}). We note that this calculation of $\mathrm{V_{max}}$ is consistent with those by \citet{novak17} and \citet{smolcic17c}.

We calculate luminosity functions, $\mathrm{\Phi(L)}$, i.e. the number of sources per comoving volume element per interval of luminosity logarithm, using:
\begin{equation}\label{eq:rlf}
$$\mathrm{\Phi(L)=\dfrac{1}{\Delta \mathrm{log(L)}}\sum_{i=1}^{N}\dfrac{1}{V_{max,i}}},$$
\end{equation}
where $\mathrm{V_{max,i}}$ is the maximum observable volume of the i-th galaxy, $\mathrm{\Delta log(L)}$ is the interval of logarithm of luminosity (here taken to be 0.8 dex), and the sum is taken over all sources within the redshift and luminosity bin under consideration. Following \citet{marshall85}, we calculate the errors on $\mathrm{\Phi(L)}$ as:
\begin{equation}\label{eq:rlf_errors}
$$\mathrm{\sigma_{\Phi(L)}=\dfrac{1}{\Delta \mathrm{log(L)}}\sqrt{\sum_{i=1}^{N}\dfrac{1}{V_{max,i}^2}}}.$$
\end{equation}
In the case where the number of sources used to calculate $\mathrm{\Phi (L)}$ in a bin is $\mathrm{\leq 10}$, we correct the values of $\mathrm{\sigma_{\Phi(L)}}$ given by Eq. \ref{eq:rlf_errors} by using the tabulated values of Poissonian uncertainty for small number statistics as given by \citet{gehrels86}.

To compute the AGN luminosity functions, we have to properly take into account the uncertainties on the AGN fractions (and therefore luminosities), which in some cases are quite significant. To do this, we use the Monte Carlo approach described in Sec. \ref{sec:decomposition} which provides us with the distribution of the AGN fraction for each object. For each Monte Carlo iteration, we split the sample into luminosity bins and calculate luminosity functions using Eq. \ref{eq:rlf}. This procedure gives us a distribution of luminosity functions per luminosity bin, from which we find the median and the corresponding 16th and 84th percentiles.

The AGN luminosity functions are tabulated in Tab. \ref{tab:lumfun} and shown in Fig. \ref{fig:lf_combined}. The median values of redshift quoted per redshift bin represent the median values of redshift of HLAGN that satisfy the flux density condition $\mathrm{S_{3\ GHz, AGN} \geq S_{lim, 3\ GHz}}$.

As expected based on comparison with the literature (e.g. \citealt{best14}, \citealt{pracy16}), we find that the total radio luminosity functions (without any decomposition applied) at high luminosities are dominated by AGN emission. At lower luminosities we can see the effect of the decomposition technique as the normalization of the AGN luminosity function is lower than that of the total luminosity function. This is expected given that at these luminosities the SFG luminosity function starts to significantly contribute to the total luminosity function (\citealt{novak18}).

%table for double column format
\begin{table*}
\begin{center}
\caption{AGN luminosity functions for the HLAGN population.}
\renewcommand{\arraystretch}{1.5}
\begin{tabular}[t]{c c c}
\hline
Redshift
& $\log\left(\dfrac{L_{1.4\,\text{GHz}}}{\text{W}\,\text{Hz}^{-1}}\right)$
& $\log\left(\dfrac{\Phi}{\text{Mpc}^{-3}\,\text{dex}^{-1}}\right)$\\
\hline 
$0.10<z<0.50$ & 21.82 &  -4.86 $\pm 0.05$ \\
$Med(z) = 0.37$ & 22.62 &   -4.74 $\pm 0.07$ \\
& 23.42 &  -5.37 $\pm 0.17$ \\
& 24.22 &  -5.28 $\pm 0.14$ \\
& 25.02 &  -5.98 $\pm 0.53$ \\
& 25.82 &  -5.98 $\pm 0.53$ \\
$0.50<z<0.70$ & 22.62 &  -4.61 $\pm 0.19$ \\
$Med(z) = 0.66$ & 23.42 &   -4.94 $\pm 0.07$ \\
& 24.22 &  -5.33 $\pm 0.12$ \\
& 25.02 &  -6.11 $\pm 0.53$ \\
$0.70<z<0.90$ & 22.96 &  -4.61 $\pm 0.09$ \\
$Med(z) = 0.82$ & 23.76 &   -4.93 $\pm 0.07$ \\
& 24.56 &  -5.41 $\pm 0.11$ \\
& 25.36 &  -6.26 $\pm 0.53$ \\
& 26.16 &  -6.26 $\pm 0.53$ \\
$0.90<z<1.10$ & 23.21 &  -5.04 $\pm 0.09$ \\
$Med(z) = 0.96$ & 24.01 &   -5.34 $\pm 0.08$ \\
& 24.81 &  -6.06 $\pm 0.30$ \\
& 25.61 &  -5.88 $\pm 0.22$ \\
& 26.41 &  -6.36 $\pm 0.53$ \\
& 27.21 &  -6.36 $\pm 0.53$ \\

\hline
\end{tabular}
\quad
\renewcommand{\arraystretch}{1.5}
\begin{tabular}[t]{c c c}
\hline
Redshift
& $\log\left(\dfrac{L_{1.4\,\text{GHz}}}{\text{W}\,\text{Hz}^{-1}}\right)$
& $\log\left(\dfrac{\Phi}{\text{Mpc}^{-3}\,\text{dex}^{-1}}\right)$\\
\hline  
$1.10<z<1.40$ &  23.42 &  -4.98 $\pm 0.08$ \\
$Med(z) = 1.20$ & 24.22 &   -5.35 $\pm 0.07$ \\
& 25.02 &  -6.13 $\pm 0.22$ \\
& 25.82 &  -6.31 $\pm 0.30$ \\
& 26.62 &  -6.13 $\pm 0.22$ \\
$1.40<z<1.70$ &  23.66 &  -5.19 $\pm 0.07$ \\
$Med(z) = 1.53$ & 24.46 &   -5.41 $\pm 0.09$ \\
& 25.26 &  -6.19 $\pm 0.22$ \\
& 26.06 &  -6.19 $\pm 0.22$ \\
$1.70<z<2.10$ &  23.85 &  -5.20 $\pm 0.07$ \\
$Med(z) = 1.91$ & 24.65 &   -5.53 $\pm 0.07$ \\
& 25.45 &  -5.94 $\pm 0.10$ \\
$2.10<z<3.00$ &  24.06 &  -5.30 $\pm 0.06$ \\
$Med(z) = 2.56$ & 24.86 &   -5.64 $\pm 0.08$ \\
& 25.66 &  -6.28 $\pm 0.10$ \\
& 26.46 &  -6.70 $\pm 0.21$ \\
& 27.26 &  -6.88 $\pm 0.30$ \\
$3.00<z<6.10$ &  24.40 &  -5.60 $\pm 0.10$ \\
$Med(z) = 3.38$ & 25.20 &   -6.31 $\pm 0.04$ \\
& 26.00 &  -6.76 $\pm 0.09$ \\

\hline
\end{tabular}
\label{tab:lumfun}
\end{center}
\end{table*}

\subsection{Comparison with the literature}\label{sec:comparison}

In Fig. \ref{fig:lf_combined} we compare our results with literature values of radio AGN luminosity functions at 1.4 GHz. 

\citet{bh12} used a combined SDSS, NVSS and FIRST sample of radio-loud AGN at redshifts $z<0.3$ to study the difference between LERGs and HERGs. For both classes they constructed local luminosity functions. Their local luminosity function of HERGs is systematically lower in comparison to our data. As previously discussed by \citet{pracy16}, this disagreement reflects the difference in the classification criteria, where \citeauthor{bh12} conservatively removed from their HERG sample all sources which showed evidence of star formation and classified them as SFGs (see also Sec. \ref{sec:llf}).

\begin{figure*}[!htbp]
  \includegraphics[width=\linewidth]{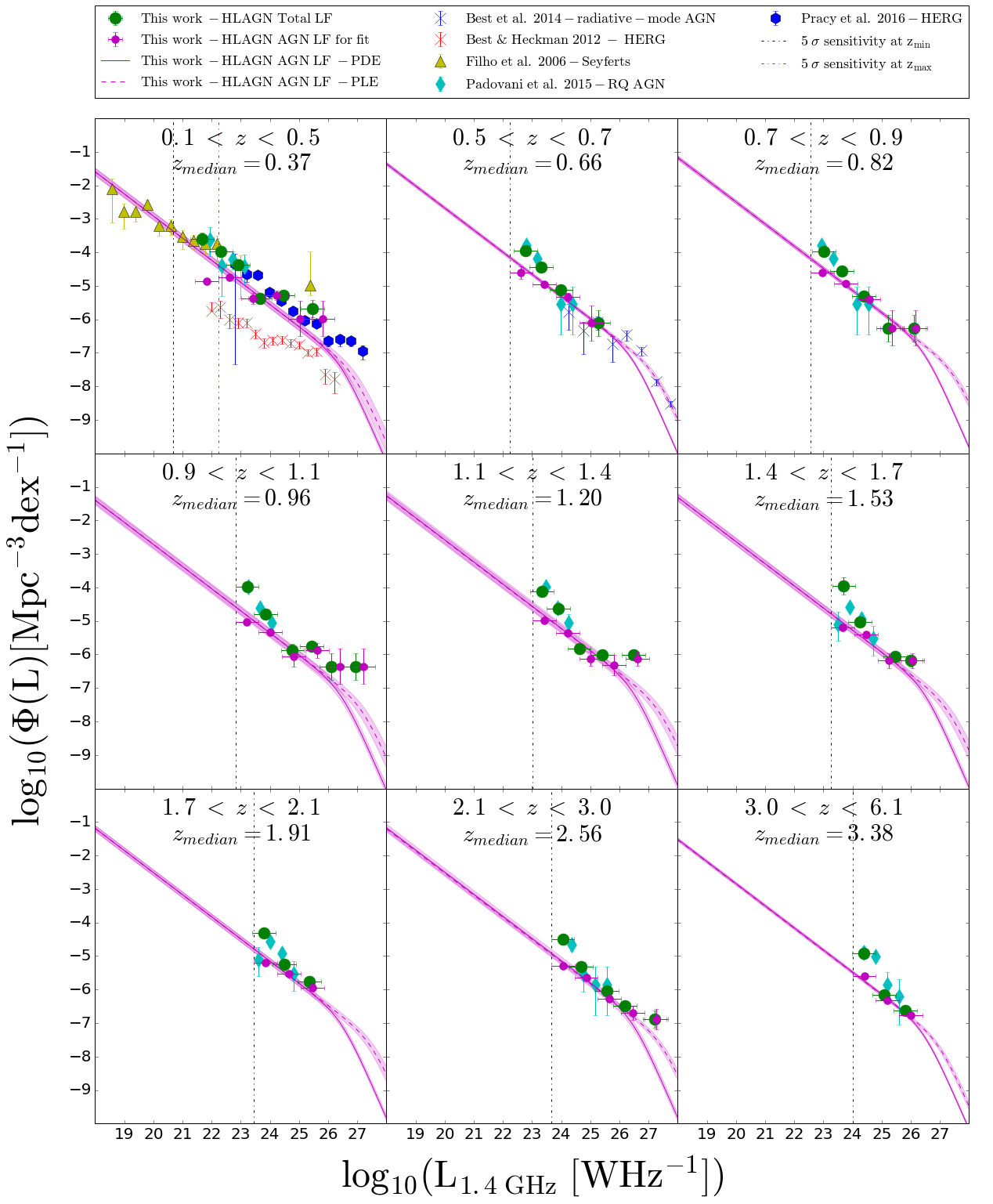}
  \caption{The radio AGN luminosity function at 1.4 GHz at redshifts $\mathrm{0.1<z<6.1}$. AGN and total luminosity functions of HLAGN are plotted as magenta and green points, respectively. Solid and dashed magenta lines show the fit of the analytic form of the local luminosity function to our AGN luminosity function data for the pure density and pure luminosity evolutions, respectively, with shaded areas indicating $\mathrm{1\sigma}$ confidence range. The black and red vertical dashed lines show the luminosity corresponding to the 5$\sigma$ sensitivity at $\mathrm{z_{min}}$ and $\mathrm{z_{max}}$ of the redshift bin, respectively. Results from the literature are also shown, as detailed in the legend.}
  \label{fig:lf_combined}\vspace{1cm}
\end{figure*}

\citet{pracy16} used a sample of radio galaxies identified by matching FIRST sources with SDSS images. Based on their optical spectra, sources were classified into LERG and HERG classes (\citealt{ching17}). In Fig. \ref{fig:lf_combined} we show their HERG luminosity functions. As will be discussed in more detail in Sec. \ref{sec:llf}, 1.4 GHz luminosity functions derived by \citeauthor{pracy16} for the HERG sample are dominated by the AGN emission and are in excellent agreement with our values of AGN luminosity functions. 

Based on the Palomar sample of bright nearby galaxies (\citealt{palomar97}) \citet{filho06} constructed 5 GHz radio luminosity functions of low-luminosity AGN at a median distance of 17 Mpc. The luminosity function of 37 Seyfert galaxies, scaled to 1.4 GHz luminosities, is in good agreement with our data and constrains the low-luminosity end of the local luminosity function.

\citet{best14} combined eight radio surveys to obtain a large source sample covering a broad range of radio luminosities. They further separated these sources into radiative-mode or jet-mode AGN based on optical emission line strengths and line flux ratios. Here we compare our data with their luminosity functions of 123 radiative-mode AGN at redshifts $0.5<z<1.0$. Their results are in agreement with our luminosity functions, with the \citeauthor{best14} luminosity functions constraining the high-luminosity end.
 
Using the VLA sample of sources within the $\sim$0.3 $\mathrm{deg^2}$ of the Extended Chandra Deep Field South (E-CDFS), \citet{padovani15} studied the evolution out to $\mathrm{z\sim 4}$ of two radio faint populations down to the sensitivity of $\mathrm{5\sigma \sim32.5\ \mathrm{\mu Jy}}$ at 1.4 GHz. To minimize the effect of missing redshift estimation, they have applied a cut in 3.6 $\mathrm{\mu m}$ flux density, $\mathrm{f_{3.6\ \mu m} > 1 \mu}$Jy, yielding a sample of 680 radio detected sources. Using the scheme described in detail in \citet{bonzini13}, these sources were classified as RL AGN, RQ AGN or SFGs. This classification was based on the logarithm of the ratio between the observed 24 $\mathrm{\mu m}$ and 1.4 GHz flux densities, $\mathrm{q_{24\ obs}}$. They define the `SFGs locus' using the SED of M82 chosen to represent the entire population of SFGs across all studied redshifts. All sources which displayed an excess of radio emission with respect to this locus were classified as RL AGN. Sources within the SFGs locus, for which X-ray and MIR diagnostics showed evidence of AGN activity, were classified as RQ AGN. The remaining sources were classified as SFGs. 

To test the evolution of their RQ AGN, \citeauthor{padovani15} split their sample in six redshift bins, up to a redshift 3.66, and within each bin calculated the 1.4 GHz luminosity functions. Their RQ AGN luminosity functions agree well with our total luminosity functions in each redshift bin, possibly due to similar classification criteria (for a detailed comparison between the classifications applied in this paper and the one used by \citet{padovani15} see \citealt{delvecchio17}). At the low luminosity end, their luminosity functions tend to be higher than our AGN luminosity functions. This is expected since the RQ luminosity functions by \citeauthor{padovani15} are based on the total 1.4 GHz luminosity which contains both star-forming and AGN contributions, while we construct HLAGN luminosity functions using only the AGN contribution to the 1.4 GHz luminosity derived with a decomposition method.

\section{Cosmic evolution of the radio luminosity function out to z$\sim$6} \label{sec:lf_evolution}

In this section we explain the reasoning behind adopting the analytic form of the local luminosity function by \citet{pracy16} as an appropriate form to fit to our data. We further describe how the evolution of radio AGN luminosity function out to $\mathrm{z\sim 6}$ is constrained.

\subsection{Local luminosity function}\label{sec:llf}

The $\mathrm{2\ deg^2}$ area of the COSMOS field does not sample a large enough volume at low redshift to constrain the local luminosity function over a wide range of radio luminosities below and above the break of the luminosity function ($\mathrm{L^{*}}$). Since an analytic representation of the local luminosity function over a broad luminosity range is required to constrain its cosmic evolution, here we consider local luminosity functions derived on the basis of larger area surveys, such as FIRST/NVSS combined with SDSS, LARGESS and the Palomar Nearby Galaxy Survey (\citealt{becker94}, \citealt{condon98}, \citealt{york00}, \citealt{ching17}, \citealt{palomar97}). In these surveys, the luminosity functions were separately derived for spectroscopically selected HERGs (\citealt{filho06}, \citealt{bh12}, \citealt{best14}, \citealt{pracy16}) which can be assumed to be the local analogs of our HLAGN (as detailed above). 
However, some significant disagreement exists between various local luminosity functions, being the strongest at the low luminosity end ($\mathrm{L_{1.4\ GHz} < 10^{25}\ \mathrm{W Hz^{-1}}}$). This disagreement is mainly due to the difference in the treatment of SFGs among different studies. \citet{bh12} took a conservative approach removing all SFGs from the sample, which resulted in the normalization of the low luminosity end being lower by an order of magnitude than that by \citet{filho06} and \citet{pracy16}. To test if the lower luminosity end of the HERG luminosity function is biased by radio emission contribution from SFGs, we independently estimate the contribution of star formation to the radio luminosity of HERGs in the LARGESS sample, used by \citet{pracy16} for the derivation of their luminosity function.

\begin{figure}[!h]
  \includegraphics[width=\linewidth]{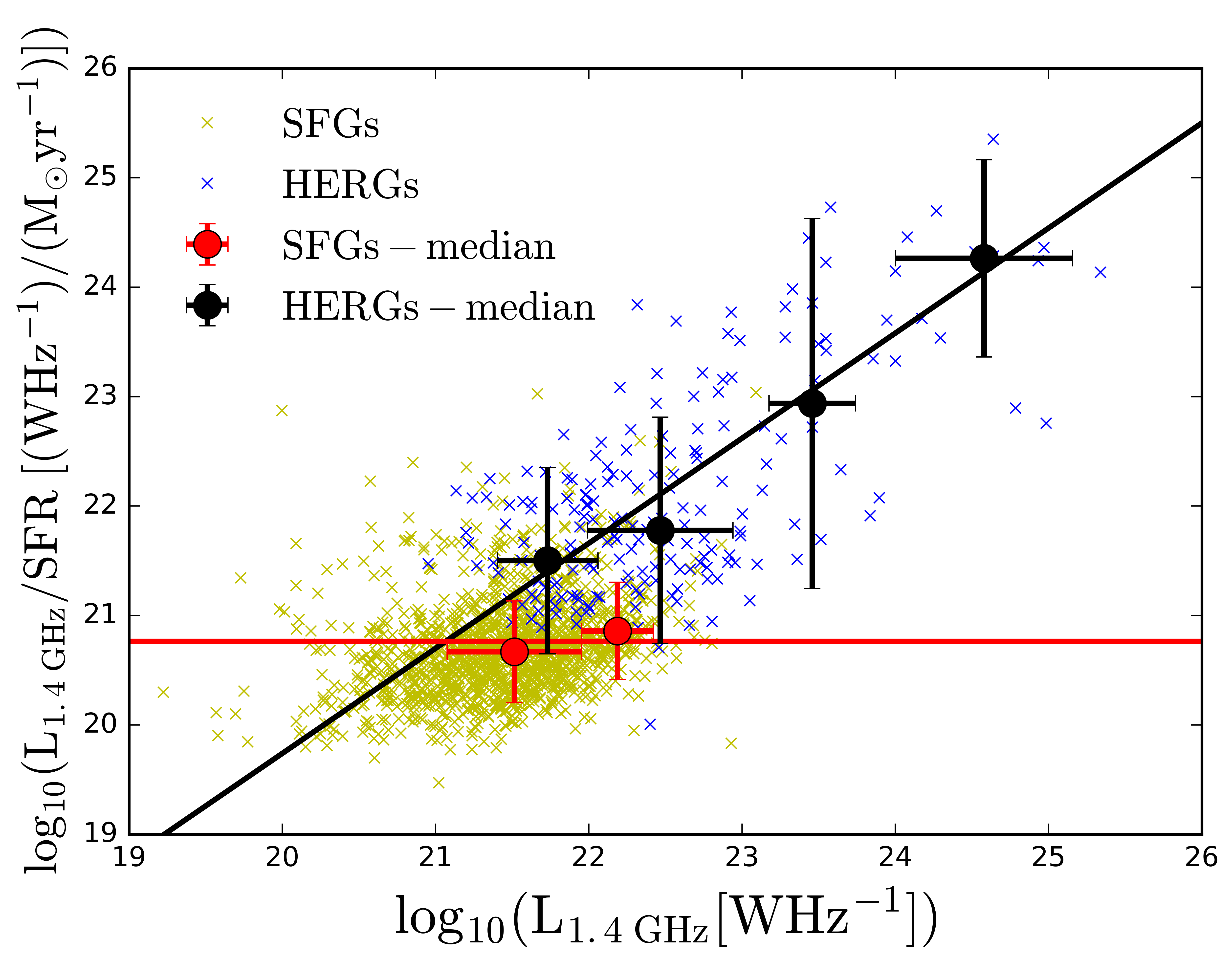}
  \caption{Ratio between the 1.4 GHz radio luminosity and SFR vs. 1.4 GHz radio luminosity for HERGs (blue symbols) and SFGs (yellow symbols) drawn from the LARGESS sample and cross-correlated with the MPA-JHU DR7 catalog. Median values of the ratio for HERGs and SFGs are shown by the filled black and red circles, respectively. The black solid line represents the linear fit to the median values for the HERG population. The red solid line represents the average value of the ratio between 1.4 GHz luminosity and SFR for star forming galaxies.}
  \label{fig:plot_lsfr_l2}
\end{figure}

We add to the LARGESS data SFR estimates provided in the MPA-JHU DR7 catalog\footnote{http://wwwmpa.mpa-garching.mpg.de/SDSS/DR7/sfrs.html}. In this catalog SFRs for sources classified as SFGs were derived directly from optical emission lines, while SFRs for AGN, Composite and Unclassified classes were estimated on the basis of the observed spectral break at  $\mathrm{4000 \AA}$ ($\mathrm{D_{4000}}$) and the relation existing for normal SFGs between specific SFR and $\mathrm{D_{4000}}$ (see \citealt{brinchmann04} for details). The given SFRs are corrected for the effect of finite SDSS spectroscopic fibers as described in \citet{salim07}.

\begin{figure}[!h]
   \includegraphics[width=\linewidth]{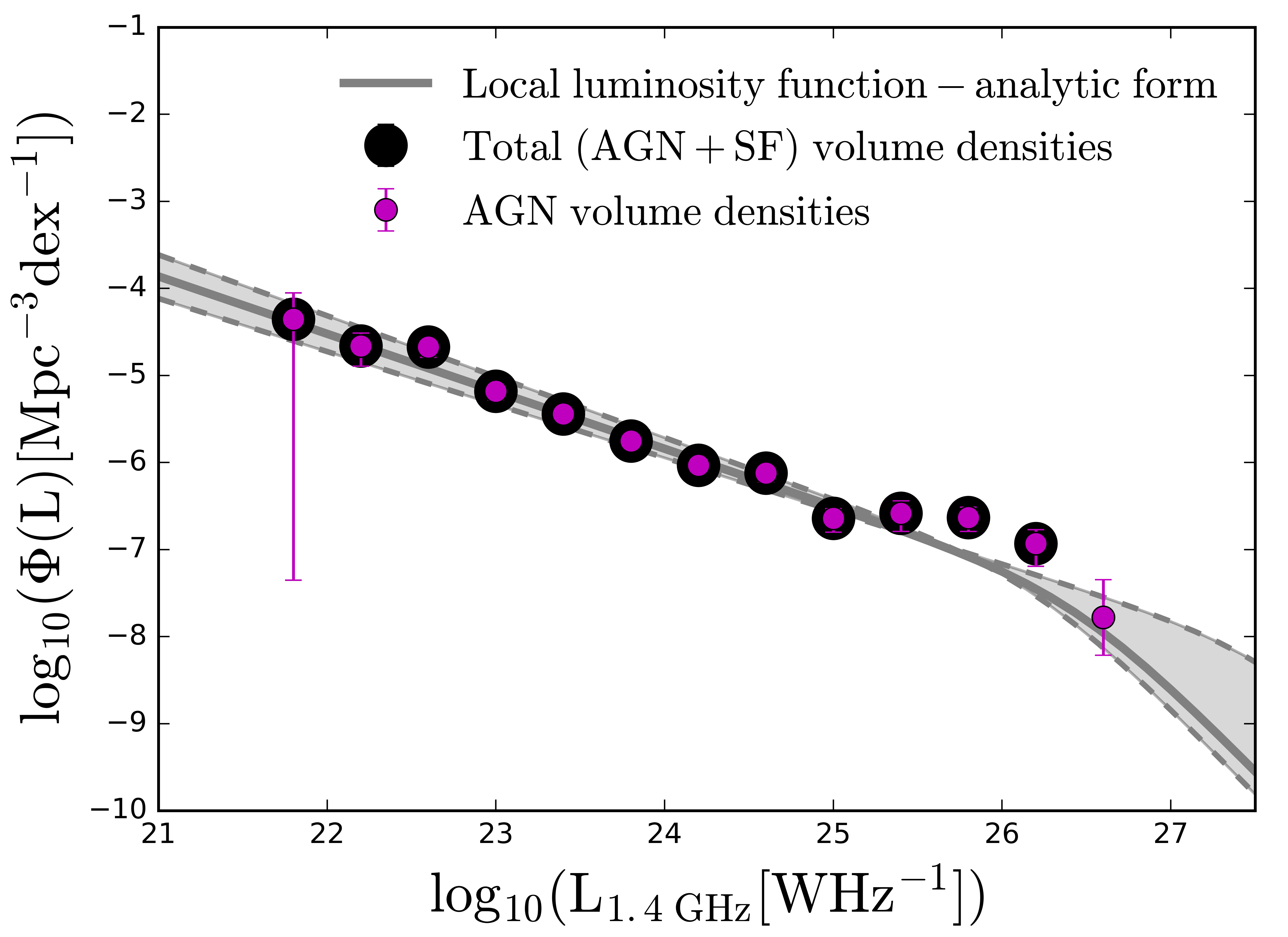}
  \caption{The local HERG luminosity function. Black circles show the total (AGN + SF) volume densities, while the gray line is the analytic fit to the data as done by \citet{pracy16}. Magenta circles show the AGN luminosity functions derived by decomposing the total luminosity. The overlap between AGN and total volume densities confirms that the radio luminosity of sources in the HERG sample by \citeauthor{pracy16} is dominated by the AGN emission.}
  \label{fig:pracy_corrected}
\end{figure}

In Fig. \ref{fig:plot_lsfr_l2} we show the ratio between 1.4 GHz luminosities and SFRs versus 1.4 GHz luminosity. As seen in this figure, the radio luminosities of HERGs are systematically higher than those of SFGs and the ratio is systematically offset from that of SFGs. By subtracting this value from the fit to the median values for the HERG population, we estimate the correlation of AGN and total 1.4 GHz luminosity of HERGs (as identified in the LARGESS sample) to be:
\begin{equation}\label{eq:lagn_pracy}
$$\mathrm{L_{AGN} = L_{TOT}(1-10^{20.22}\  L_{TOT}^{-0.96})}$$. 
\end{equation}
We use relation (\ref{eq:lagn_pracy}) to statistically decompose the total radio luminosities of the HERG sample provided in Table 1 in \citeauthor{pracy16} into the luminosity components originating from an AGN and star-forming processes and then we compute the luminosity function using these estimates for the AGN radio luminosities. The results are plotted in Fig. \ref{fig:pracy_corrected} where we compare them to the original luminosity functions by \citet{pracy16}. The decomposed radio luminosity function, as derived here, is perfectly consistent with that presented by \citet{pracy16}. From this we conclude that the total luminosity of sources in the HERG sample is dominated by the AGN emission. The analytic form of the local HERG luminosity function as derived by \citeauthor{pracy16} can hence be further used to constrain the cosmic evolution of our HLAGN sample (which is selected to trace the high redshift analogs of HERGs).

\subsection{Cosmic evolution of the radio AGN luminosity function} \label{sec:lf_evolution_sub}

The purpose of deriving the radio AGN luminosity functions of the HLAGN sample in Sec. \ref{sec:lf_derivation} is to study their evolution over cosmic time. We test their evolution with the pure density evolution (PDE) and pure luminosity evolution (PLE) models. We fit the analytic form of the local luminosity function to our data: 
\begin{equation}\label{eq:rlf_evolution}
$$\mathrm{\Phi(L,z)=(1+z)^{\alpha_{D}}\ \Phi_{0}\left[ \dfrac{L}{(1+z)^{\alpha_{L}}}\right], }$$
\end{equation}
where $\mathrm{\alpha_{D}}$ and $\mathrm{\alpha_{L}}$ are pure density and pure luminosity evolution parameters, respectively. $\mathrm{\Phi_{0}}$ is the local luminosity function as defined by \citet{pracy16}:
\begin{equation}\label{eq:lf_pracy}
$$\mathrm{\Phi_{0} (L)  = \dfrac{\Phi^{*}}{\left(L^{*}/L\right)^{\alpha} + \left(L^{*}/L\right)^{\beta}}},$$
\end{equation}
where $\mathrm{\Phi^{*} = 10^{-7.47} \ \mathrm{Mpc^{-3} dex^{-1}}}$ is the normalization, $\mathrm{L^{*} = 10^{26.47} \ \mathrm{WHz^{-1}}}$ is the knee of the luminosity function and $\mathrm{\alpha = -0.66}$ and $\mathrm{\beta < 0}$ are the faint and bright end slopes, respectively. This analytic form was constrained using the local ($\mathrm{0.005<z<0.3}$) sample of HERGs down to the optical apparent magnitude limit of their survey ($\mathrm{m_i < 20.5}$). As discussed by \citeauthor{pracy16} (\citeyear{pracy16}; see Sec. 3.6), due to the poor statistics they give only an upper limit for the parameter $\mathrm{\beta}$. In our analysis, we set the bright end slope to the value of -2 as it matches well the radiative-mode AGN luminosity functions derived by \citeauthor{best14} (\citeyear{best14}; see Fig.\ref{fig:lf_combined}). However, to properly constrain this parameter, better statistics of bright objects are needed.

\begin{figure}[h]
  \includegraphics[width=\linewidth]{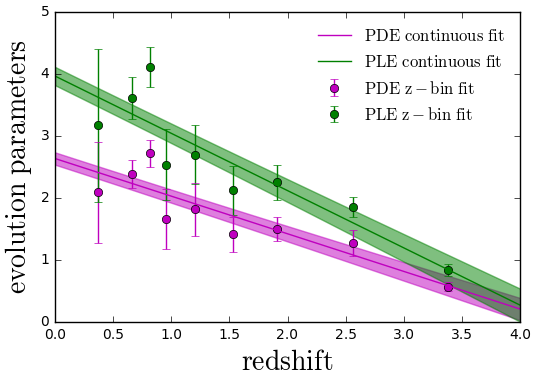}
  \caption{Parameters obtained from fitting two different redshift evolution models to the HLAGN luminosity functions. Green and magenta circles show the evolution parameters obtained from fitting the assumed analytic form of the luminosity function in nine redshift bins assuming pure luminosity and pure density evolution scenarios, respectively (see text for details). The same color lines show the results from the continuous fit assuming that the PLE and PDE parameters evolve linearly with redshift.}
  \label{fig:hlagn_evolution_parameters}
\end{figure}

To constrain the evolution of the HLAGN sample, we use the AGN luminosity function data with luminosities above the $\mathrm{5\sigma}$ luminosity threshold of our survey at the minimum redshift of the bin ($\mathrm{L(z_{min})}$). We make an exception in the first redshift bin ($\mathrm{0.1<z<0.5}$), where our lower luminosity data are not well constrained due to the small area of the COSMOS field. In this bin, we fit the analytic form of the local luminosity function by \citeauthor{pracy16} only to the data which are above the luminosity threshold of our survey at the maximum redshift of the bin ($\mathrm{L(z_{max})}$).

Firstly, we fit the analytic form of the local luminosity function to the data in nine redshift bins. We test the PDE and PLE models by setting either $\mathrm{\alpha_{D} = 0}$ or $\mathrm{\alpha_{L}=0}$ for the case of pure luminosity or density evolution, respectively. Evolution parameters for each redshift bin obtained with this procedure are presented in Table \ref{tab:alpha_evol} and Fig. \ref{fig:hlagn_evolution_parameters}.

\begin{table}
\begin{center}
\caption{Best-fit evolution parameters obtained by the fitting local luminosity function to the redshift binned data assuming pure density ($\alpha_D$) and pure luminosity ($\alpha_L$) evolution.}
\renewcommand{\arraystretch}{1.5}
\begin{tabular}[t]{c c c}
\hline
Med(z)
& $\alpha_{D}$
& $\alpha_{L}$\\
\hline 
0.37 & 2.09 $\pm 0.82$ &  3.17 $\pm 1.23$ \\
0.66 & 2.38 $\pm 0.22$ &  3.61 $\pm 0.33$ \\
0.82 & 2.72 $\pm 0.22$ &  4.11 $\pm 0.32$ \\
0.96 & 1.66 $\pm 0.48$ &  2.54 $\pm 0.57$ \\
1.20 & 1.82 $\pm 0.43$ &  2.70 $\pm 0.48$ \\
1.53 & 1.41 $\pm 0.28$ &  2.12 $\pm 0.39$ \\
1.91 & 1.50 $\pm 0.19$ &  2.26 $\pm 0.28$ \\
2.56 & 1.28 $\pm 0.21$ &  1.86 $\pm 0.16$ \\
3.38 & 0.57 $\pm 0.08$ &  0.83 $\pm 0.10$ \\

\hline
\end{tabular}
\quad
\label{tab:alpha_evol}
\end{center}
\end{table}

Secondly, we continuously model the redshift dependence of the evolution parameters following the procedure described by \citet{novak17}. We fit a simple linear redshift-dependent evolution model to all AGN luminosity functions in all redshift bins simultaneously:
\begin{equation}\label{eq:continuous_evolution}
$$\mathrm{\Phi(L,z)=(1+z)^{\alpha_d+z\cdot\beta_d}\times\Phi_0 \left[ \dfrac{L}{(1+z)^{\alpha_l+z\cdot\beta_l}} \right]}, $$
\end{equation}
where $\mathrm{\alpha_d}$, $\mathrm{\alpha_l}$, $\mathrm{\beta_d}$ and $\mathrm{\beta_l}$ are the various evolution parameters. We again test pure density and pure luminosity evolutions separately via a $\mathrm{\chi^2}$ minimization procedure. The best-fit parameters obtained in the case of pure luminosity evolution ($\mathrm{\alpha_d=\beta_d =0}$) are $\mathrm{\alpha_l = 3.97\pm 0.15}$ and $\mathrm{\beta_l=-0.92\pm 0.06}$. In the case of pure density evolution ($\mathrm{\alpha_l=\beta_l =0}$) the best-fit parameters are $\mathrm{\alpha_d = 2.64\pm 0.10}$ and $\mathrm{\beta_d=-0.61\pm 0.04}$. The values of these continuous evolution parameters and their uncertainties are shown with color lines in Fig. \ref{fig:hlagn_evolution_parameters}.

Our results show a global decline of evolution parameters with redshift. The high parameters of evolution in the redshift bin $\mathrm{0.7<z<0.9}$, significantly higher than those derived with the linear redshift dependent evolution model, are likely due to the presence of a very prominent large scale structure in the COSMOS field at $\mathrm{z\sim0.73}$ (\citealt{guzzo07}, \citealt{iovino16}). The presence of such a structure, enhancing the normalization of the luminosity function, produces higher than expected parameters of evolution.

To test if the chosen redshift binning has a significant effect on our results, we repeat the entire analysis using the binning designed to contain roughly the same number of sources with $\mathrm{S_{3\ GHz, AGN}\geq S_{lim, 3\ GHz}}$ per redshift bin. Results of the best-fit parameters of a simple linear redshift dependent evolution model obtained through this analysis  are consistent within the confidence range ($\mathrm{\pm 1\sigma}$) with the results presented above.

\section{Discussion} \label{sec:discussion}

In this section we discuss our results in the context of galaxy evolution, also investigating the origin of the radio emission in HLAGN. We estimate the evolution of number, luminosity and kinetic luminosity density and compare our results to other studies of galaxy evolution.

\subsection{Origin of radio emission in HLAGN} \label{sec:radio_origin}

Previous studies conducted on radio samples of X-ray and mid-IR selected AGN suggest that AGN identified through these criteria are hosted mostly by star-forming systems in which the bulk of the radio emission originates from the host galaxy rather than SMBH activity (e.g. \citealt{hickox09}, \citealt{bonzini13}, \citealt{goulding14}, \citealt{padovani15}). 
For instance, in a comprehensive analysis performed by \citet{delvecchio17} on the HLAGN sample they report that only $\sim$30$\%$ of them display significant radio excess due to non-thermal emission originating from AGN, while the radio emission of the rest is dominated by star formation. However, this behavior does not apply to the most luminous radio sources, which are found to mostly reside in massive and passive systems (e.g. \citealt{dunlop03}).

In this work we have separated the radio emission of HLAGN into radiation originating from star formation and AGN activity using a statistical decomposition technique. By deriving AGN fractions for all galaxies within our sample, we were able to estimate the extent to which the AGN component contributes to the total radio luminosity.
Based on our estimates of AGN fractions we found that the majority, $\mathrm{(68.0\pm 1.5)\%}$, of the HLAGN are dominated by star-forming processes ($\mathrm{0\leq f_{AGN}\leq 0.5}$), while in $\mathrm{(32.0\pm1.5)\%}$ of HLAGN AGN activity dominates ($\mathrm{0.5 <f_{AGN} \leq 1.0}$) in the radio regime (see Sec. \ref{sec:decomposition}). We found no significant redshift evolution of the AGN fraction for HLAGN with AGN-related radio emission.

\begin{figure}[!h]
 \includegraphics[width=\linewidth]{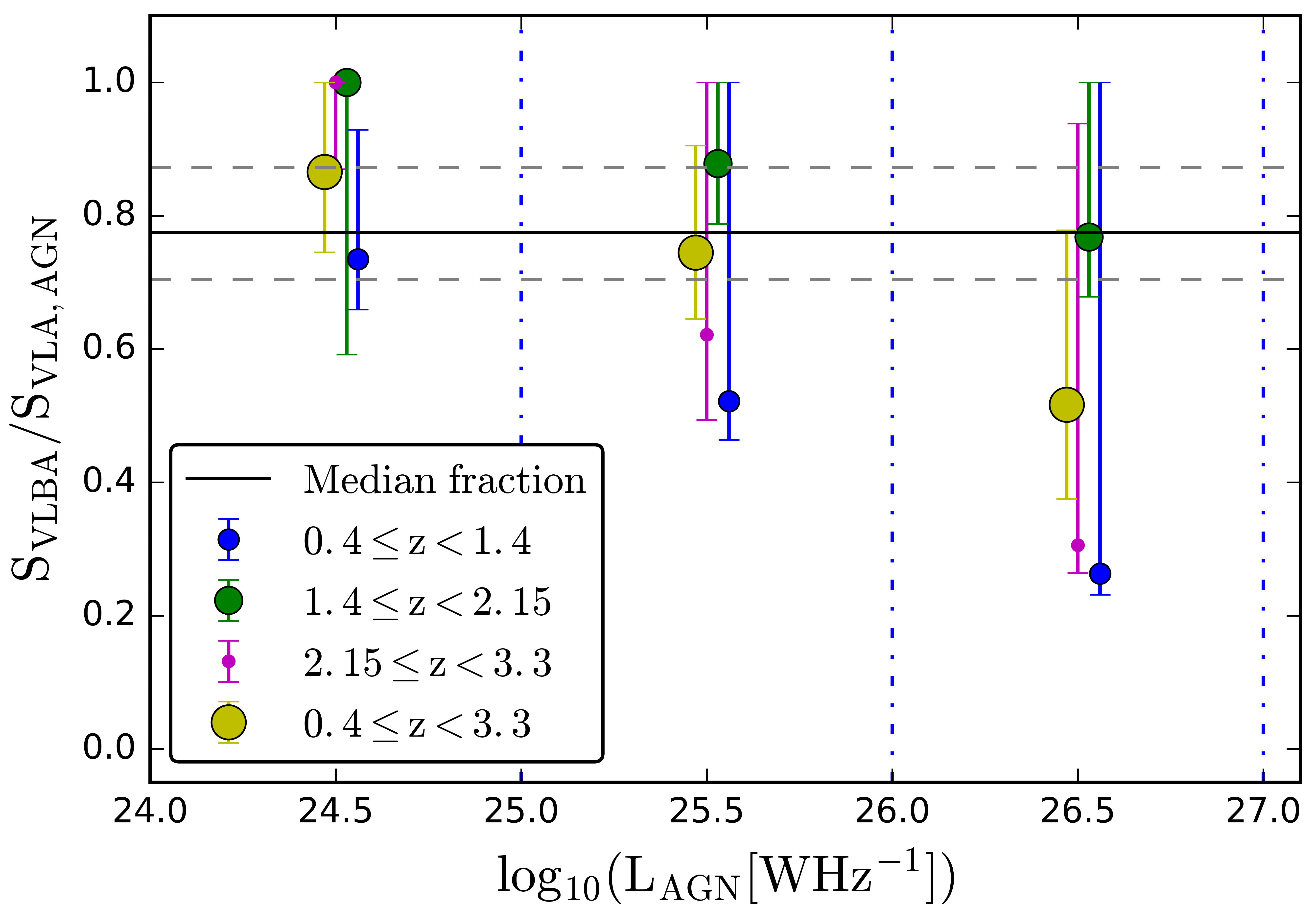}
  \caption{The VLBA-to-VLA AGN flux density ratio vs. 1.4 GHz AGN luminosity. Yellow circles show the luminosity binned values of the ratio, while solid and dashed black lines show the median, 84th and 16th percentile values of the sample with $\mathrm{L_{\mathrm{1.4\ GHz, TOT}}\geq 10^{24.4}\ \mathrm{WHz^{-1}}}$. Blue, green and magenta circles and error bars show the data in three redshift bins as indicated in the legend. For better visibility of error bars, we introduced a small shift along the luminosity axis.}
  \label{fig:vla_vlba_flux}
\end{figure}

The question that arises next is whether radio AGN emission is extended or confined to the galaxies' nuclear regions. To answer this question, we cross-match our 3 GHz sample of 1,604 HLAGN with the Very Long Baseline Array (VLBA) 1.4 GHz radio catalog by \citet{herreraruiz17}. 
The authors identified a sample of 468 radio AGN detected by VLBA previously targeted with the VLA-COSMOS survey at 1.4 GHz (\citealt{schinnerer10}). 
They found that roughly $\sim$70$\%$ (237/344) of the faint sources ($\mathrm{S_{VLA} \leq 1 mJy}$) are dominated by emission originating from a compact core, while this is true for only $\sim$30$\%$ (6/21) of the bright population ($\mathrm{S_{VLA} >10\ mJy}$). They argue that the bulk of the radio emission of the bright population originates from structures larger than the VLBA scale.
To check whether these conclusions hold also for HLAGN, we limit our HLAGN sample to match the original VLBA selection. 

To match the VLBA selection, we require a $\mathrm{>5.5\sigma}$ VLA detection at 1.4 GHz\footnote{The average rms reported by \citet{schinnerer10} is 12 $\mathrm{\mu Jy}$/beam.}. Out of 619 HLAGN satisfying this condition, we find only 113 ($\mathrm{18\%}$) to have a VLBA counterpart within a 0.4" radius corresponding to about half a beam size of the 3 GHz VLA-COSMOS Large Project observations. 
The remaining 82$\%$ of HLAGN were not detected with the VLBA either because they were not detected within the VLA-COSMOS survey at 1.4 GHz ($\mathrm{22\%}$) or due to their radio emission being extended to scales larger than those probed with the VLBI method\footnote{Physical scales of VLBA sources range from $20\ \mathrm{pc} $ at $z=0.1$ to $80\ \mathrm{pc} $ at $z\sim3$}($\mathrm{60\%}$). 

To study the properties of the VLBA detected HLAGN, we restrict our analysis to a luminosity-limited subsample of sources with 1.4 GHz luminosities $\mathrm{L_{\mathrm{1.4\ GHz}}\geq 10^{24.4}\ \mathrm{WHz^{-1}}}$ (the lowest 1.4 GHz luminosity of the sources at redshifts $\mathrm{z\geq2}$). This criterion yields 51 ($\mathrm{8\%}$) HLAGN, spanning redshifts of $\mathrm{0.4<z<3.3}$. By scaling the 1.4 GHz VLA fluxes by AGN fractions, we estimate the flux expected to originate from the AGN component of these sources (see Sec. \ref{sec:decomposition}). Using these fluxes we further calculate the VLBA-to-VLA AGN flux density ratios. To find the median value and the corresponding $\mathrm{1\sigma}$ confidence range of the flux density ratios, we perform Monte Carlo simulations taking into account the measurement errors of the 1.4 GHz VLA and VLBA flux densities and AGN fractions. 

For the sample of the VLBA detected HLAGN with $\mathrm{L_{\mathrm{1.4\ GHz}}\geq10^{24.4}\ \mathrm{WHz^{-1}}}$, we find the median VLBA-to-VLA AGN flux density ratio to be $0.78^{+0.09}_{-0.07}$. By separating the sample in bins of the AGN luminosity ($\mathrm{\Delta \mathrm{log} L_{AGN}=1}$), we find that the higher AGN luminosity sources tend to have lower VLBA-to-VLA AGN flux density ratios (see Fig. \ref{fig:vla_vlba_flux}). These results suggest that the most powerful AGN are dominated by more extended radio emission. 
To test whether this trend changes with redshift, we separate the sample into three redshift intervals with equal number of sources (17 per bin). A trend of higher AGN luminosity sources having lower flux density ratios is present at all redshifts. Median flux density ratios of the sources with $\mathrm{L_\mathrm{1.4\ GHz, AGN} > 10^{26}\ \mathrm{WHz^{-1}}}$ at redshifts $\mathrm{z<1.4}$ and $\mathrm{2.15 <z<3.3}$ hint that over $\mathrm{50 \%}$ of the AGN related radio emission might arise from more extended structures (relative to scales probed by the VLBA). This is in agreement with the above mentioned results by \citet{herreraruiz17}.

The physical explanation for the variety of the observed radio AGN fractions in our HLAGN sample is most likely closely related to the intrinsic properties of individual AGN: the mass and spin of the central SMBH and accretion rate (e.g. \citealt{king08}). \citet{fanidakis11} used GALFORM, the semi-analytical simulation of formation and evolution of galaxies, to study the effect these properties might have on the appearance of an AGN. Their simulation shows that massive, rapidly spinning BHs are hosted by giant early-type galaxies, while lower mass BHs with much lower spins are mostly found in late-type galaxies. On the basis of the Blandford-Znajeck (BZ) mechanism (\citealt{bz77}), the authors suggest radio-loudness to be determined by the spin of the BH as only a rotating SMBH can form radio jets. In this scenario, the range of AGN fractions in our HLAGN sample suggests they may have a range of BH spins, ranging from AGN with rapidly rotating BHs with prominent AGN emission to BHs with low spin in which faint AGN emission might be diluted by the light of the host galaxy. This is supported by the results from \citet{fb99} who derived equations for a scaled down AGN jet model based on equipartition assumptions and applied it to observationally well defined samples of galactic and extragalactic sources. They found that the jet powers of their studied sources were comparable to their accretion disk luminosities providing evidence for a disk-jet coupling holding from stellar mass BHs to low-luminosity AGN.

\citet{mh08} presented a synthesis model for the AGN evolution where they studied how SMBHs grow and evolve through the age of the Universe. In this model, AGN with high Eddington ratios may display either only radiative feedback (high-radiative mode, HR) or both radiative and kinetic feedback (high-kinetic mode, HK). In this context, HLAGN with low radio AGN fractions $\mathrm{(f_{AGN} < 0.5)}$ would be the ones which display mostly radiative feedback, while those with high radio AGN fractions  $\mathrm{(f_{AGN} \geq 0.5)}$ produce also kinetic feedback in the form of radio-detectable features, such as cores, jets and lobes.

Radiative feedback is expected to arise in galaxies hosting rapidly accreting SMBHs, where the energy released by the growth of the SMBH can influence the properties of the hosts (e.g. see a review by \citealt{fabian12}). The models of galaxy evolution usually find the AGN bolometric luminosity density to increase from the local redshift out to a peak around $z\sim 2-3$, followed by a decrease towards higher redshifts (e.g. \citealt{croton06}, \citealt{mh08}). For the $\sim38\%$ of HLAGN for which we estimated a radio luminosity arising from AGN emission (Sec. \ref{sec:decomposition}) we compare the radiative luminosity density to literature expectations (e.g. \citealt{croton06}, \citealt{mh08}). We bin the radiative luminosities derived from the SED-fitting decomposition (see \citealt{delvecchio17}) into nine redshift bins, finding typical values of bolometric luminosity density to be $<1 \%$ of those found from the literature for all radiatively efficient AGN. This confirms that X-ray and MIR selected AGN detected in radio are not the dominant contribution to the total bolometric output expected for radiatively efficient AGN.

In the following sections we describe how the number and luminosity densities of HLAGN change through cosmic time. By scaling the 1.4 GHz AGN luminosity into kinetic luminosity, we calculate the kinetic luminosity density and compare it to semi-analytic model predictions. 

\subsection{The cosmic evolution of radio AGN} \label{sec:nd_ld}

To calculate the number and luminosity density, we use:
\begin{equation}\label{eq:number_density}
$$\mathrm{\mathscr{N} = \int \Phi (L_{1.4\ GHz})\ d(\mathrm{log}L_{1.4\ GHz}),}$$
\end{equation}
\begin{equation}\label{eq:luminosity_density}
$$\mathrm{\mathscr{L} = \int L_{1.4\ GHz}\times \Phi (L_{1.4\ GHz})\ d(\mathrm{log}L_{1.4\ GHz}),}$$
\end{equation}
where $\mathrm{\Phi (L_{1.4\ GHz})}$ is the redshift evolved luminosity function, using both best-fit evolution parameters derived from the redshift binned and continuous fitting. Here we take the lower and upper limits of the integral to be $\mathrm{L_{1.4\ GHz} = 10^{22} \ \mathrm{WHz^{-1}}}$ and $\mathrm{L_{1.4\ GHz} = 10^{28} \ \mathrm{WHz^{-1}}}$, respectively \footnote{The normalizations of the number and luminosity density evolution curves are affected by the chosen limits of integration. For example, a decrease of the lower limit to $\mathrm{L_{1.4\ GHz} = 10^{21} \ \mathrm{WHz^{-1}}}$ would produce a number density higher by a factor of $\sim 5$, but it would have almost no effect on the luminosity density curve. However, a decrease of the upper limit to $\mathrm{L_{1.4\ GHz} = 10^{27} \ \mathrm{WHz^{-1}}}$ would lower the normalization of the luminosity density by a factor of $\mathrm{\sim 2}$, while the number density would remain unchanged.}.

\begin{figure}[!h]
 \includegraphics[width=\linewidth]{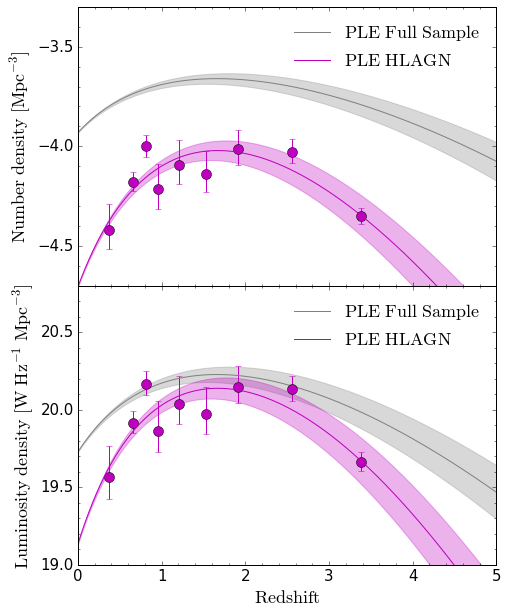}
  \caption{Redshift evolution of the number and luminosity density of AGN in COSMOS field for the PLE model are shown in the upper and lower panels, respectively. Magenta circles show the values of number and luminosity density for the HLAGN sample calculated from the redshift binned evolution parameters. Magenta and gray lines show the continuous redshift evolution and corresponding $1\sigma$ confidence range of HLAGN and the full sample, respectively.}
  \label{fig:densities}
\end{figure}

The evolution of the number and luminosity densities is shown in Fig. \ref{fig:densities}. We find that the number density of HLAGN increased over time, reaching a maximum of $\sim 9.5 \times 10^{-5}\  \mathrm{Mpc^{-3}}$ in the redshift range $1<z<2.5$, after which it decreased by a factor of $\sim$5 toward the local ($z=0$) value. A similar trend is seen in the luminosity density evolution, with the peak at $\sim 1.4\times10^{20}\ \mathrm{WHz^{-1}Mpc^{-3}}$ occurring in the redshift range $1<z<2.5$ followed by a decline of a factor $\sim$10 toward the local value. 
An outlier in redshift bin $\mathrm{0.7<z<0.9}$ seen in Fig. \ref{fig:densities} most likely occurs due to a large scale structure known to exists in the COSMOS field at $\mathrm{z\sim 0.73}$, as previously discussed in Sec. \ref{sec:lf_evolution_sub}.

In Fig. \ref{fig:densities} we also compare our results for the HLAGN sample with the results derived for all sources containing an AGN-related radio emission in the VLA-COSMOS 3 GHz Large Project sample with COSMOS2015 counterparts (labeled as `full sample'). The analysis of the full sample is described in Appendix \ref{sec:comp_c15}. The number\footnote{\citet{ms07} local luminosity function used to constrain the evolution of the full sample has a different shape than the one by \citet{pracy16} used to constrain the evolution of HLAGN. The difference between the two local luminosity functions results in a higher normalization of the number density curve for the full sample with respect to the HLAGN curve.} and luminosity densities of the full sample flatten over the redshift range $1<z<2.5$ at $\sim 2.2 \times 10^{-4}\  \mathrm{Mpc^{-3}}$  and $1.7\times10^{20}\ \mathrm{WHz^{-1}Mpc^{-3}}$, respectively. 
The more rapid decrease with redshift of the HLAGN number and luminosity density at $z>2$ when compared to the full sample may be in part explained by the incompleteness of X-ray data used to classify sources as HLAGN (see also Discussion in \citealt{smolcic17c}).

\subsection{Kinetic feedback of radio AGN and comparison with semi-analytic models} \label{sec:kinetic_feedback}

To get an insight into how the kinetic feedback of AGN changes through cosmic time, one of many scaling relations available in the literature (e.g. \citealt{willott99}, \citealt{mh07}, \citealt{gs16}) can be used to estimate the kinetic luminosity (see, e.g., Appendix A in \citealt{smolcic17c}). \citet{willott99} used the minimum energy condition to estimate the energy stored in lobes from the observed monochromatic synchrotron emission, finding the relation:
\begin{equation}\label{eq:willott}
$$\mathrm{log(L_{kin})=0.86\ logL_{1.4GHz}+14.08+1.5\ logf_{W},}$$
\end{equation}
where $\mathrm{L_{kin}}$ is the kinetic luminosity, $\mathrm{L_{1.4\ GHz}}$ is the $\mathrm{1.4\ GHz}$ radio luminosity and $\mathrm{f_{W}}$ is the factor into which all the uncertainties were combined. Using observational constraints, $\mathrm{f_{W}}$ is estimated to be in the range of $1 - 20$. In this work, the \citeauthor{willott99} conversion is used to estimate the kinetic luminosity of HLAGN and the full radio detected samples. We assume $\mathrm{f_W = 15}$ for which the \citeauthor{willott99} relation agrees with results found from radio and X-ray observations of cavities in galaxy groups and clusters (\citealt{birzan04}, \citealt{o'sullivan11}).

\begin{figure}[!h]
  \includegraphics*[width=\linewidth]{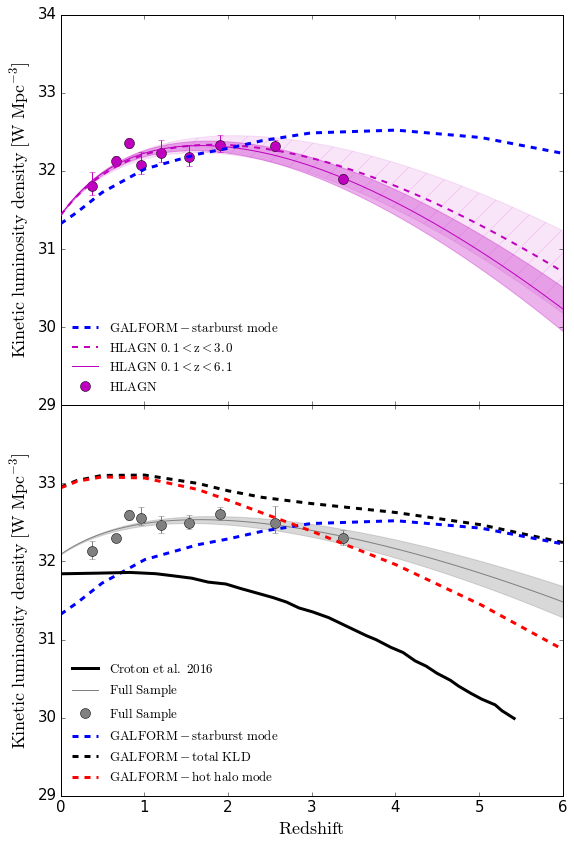}
	\caption{Cosmic evolution of the kinetic luminosity density. In the upper panel, the kinetic luminosity density derived using the \citet{willott99} conversion between radio and kinetic luminosity for the PLE continuous and redshift binned model of HLAGN is shown with a magenta line and circles, respectively, with the corresponding $\mathrm{1\sigma} $ uncertainties. The KLD estimate for the starburst mode of accretion based on the semi-analytic GALFORM model as calculated by Griffin et al. (in prep.) is shown with a blue dashed line. Lower panel shows the KLD for the PLE continuous and redshift binned models of the full sample (gray line and circles, respectively), with corresponding $\mathrm{1\sigma}$ uncertainties. The black solid line shows the prediction of the semi-analytical model for the radio mode of BH accretion by \citet{croton16}. Estimates of the KLD based on the GALFORM model for the starburst and hot-halo modes of accretion and their sum are shown with blue, red and black dashed lines, respectively.}
  \label{fig:mech_power}
\end{figure}

To calculate the kinetic luminosity density at a given redshift, we use the relation:
\begin{equation}\label{kin_pow}
$$\mathrm{\Omega_{kin} (z) = \int L_{kin} (L_{1.4\ GHz}) \times \Phi (L_{1.4\ GHz})\ d(logL_{1.4\ GHz}),}$$
\end{equation}
where $\mathrm{\Phi (L_{1.4\ GHz})}$ is the analytic form of the local luminosity function (see Eq. \ref{eq:lf_pracy}) and $\mathrm{\Omega_{kin}}$ is the kinetic luminosity density (KLD; see Fig. \ref{fig:mech_power}). The normalization of the KLD curve is highly affected by the \citet{willott99} uncertainty factor $\mathrm{f_W}$, whose values can shift the curve both towards lower and higher values\footnote{The kinetic luminosity density (KLD) can be scaled from factor $\mathrm{f_{W}^{1}}$ to $\mathrm{f_{W}^{2}}$ using relation: 
$$\mathrm{KLD(f_{W}^{2}) = KLD(f_{W}^{1}) \left( \dfrac{f_{W}^{2}}{f_{W}^{1}}\right)^{3/2}}.$$
The ratio of the $\mathrm{KLD\ (f_{W}=20)}$ and $\mathrm{KLD\ (f_{W}=1)}$ is $\mathrm{\sim 90}$, i.e. the maximum difference in estimates of the KLD using $\mathrm{f_{W}}$ factors is almost two orders of magnitude.}. By assuming $\mathrm{f_W=15}$, the KLD curve of HLAGN peaks at the value $\mathrm{2.1 \times 10^{32}\ \mathrm{WMpc^{-3}}}$ at $\mathrm{z\sim 1.5}$. The full sample KLD curve is relatively flat in the redshift range $\mathrm{1<z<2.5}$ at the value $\mathrm{3.4 \times 10^{32}\ WMpc^{-3}}$.

Using the calculations embedded in the GALFORM semi-analytic model of formation and evolution of galaxies (\citealt{bower06}), \citet{fanidakis12} studied the evolution of AGN across cosmic time. In their model, the buildup of BH mass occurs in two different modes: (i) starburst mode in which the accretion event is triggered by disk instabilities or galaxy mergers, and (ii) hot-halo mode in which the gas from a quasi-hydrostatic hot halo accretes onto the SMBH. They find that different accretion channels show different evolutionary trends, with starburst and hot-halo modes changing their dominance in the buildup of SMBH mass from higher to lower redshifts, respectively. An update of the \citeauthor{fanidakis12} model, using a new version of GALFORM (\citealt{lacey16}), will be presented in Griffin et al. (in prep.). After the implementation of the Planck cosmology and using a higher resolution dark matter simulation, one of their predictions is the cosmic evolution of kinetic luminosity density for the starburst and hot-halo modes of BH accretion (see Fig. \ref{fig:mech_power}). The kinetic luminosity density is expected to arise from AGN jets in both of these modes. In the hot halo mode gas from the hot halo slowly accretes onto the SMBH, forming an advection dominated accretion flow (ADAF; \citealt{ny94}). On the other hand, in the starburst mode a merger or a disk instability causes stars to form from inflowing gas which may also accrete onto the SMBH, in the form of a standard thin disk accretion (\citealt{ss73}). 

The selection of our HLAGN sample based on X-ray and MIR criteria aims to trace the thin disk accreters and therefore should be compared to the starburst mode AGN from GALFORM. In the upper panel of Fig. \ref{fig:mech_power}, the HLAGN KLD curve shows a steep increasing trend similar to that predicted for the starburst mode of accretion out to a $\mathrm{z\sim2}$. At higher redshifts the HLAGN evolutionary curve disagrees with predictions for the starburst mode of accretion. A possible reason for the disagreement could be an overly simplistic conversion between the monocromatic and kinetic luminosity. A more complex conversion dependent on various parameters, such as synchrotron age and environments (\citealt{hardcastle14}, \citealt{kapinska14}), could alter the overall shape of the KLD curve. Another reason could be the incompleteness of the HLAGN sample at high redshifts (\citealt{delvecchio17}). If we were to exclude the highest redshift measurement and fit the evolution to the data, our results would show agreement to GALFORM prediction for the starburst mode out to a $\mathrm{z\sim3}$ (as shown in the upper panel of Fig. \ref{fig:mech_power}). However, the discrepancy still remains between our results and the GALFORM prediction present at higher redshift. We note that, as detailed below, various semi-analytic models tend to disagree in this regime as well. 

In the lower panel of Fig. \ref{fig:mech_power} we show the comparison between the KLD curve derived for the full sample and the prediction from semi-analytic models. Despite the difference in normalizations, the KLD curve of the full sample interestingly shows a similar shape to the total KLD estimate from GALFORM, both of which contain kinetic luminosity contribution of all AGN, regardless of the assumed mode of accretion. 

We also compare our results of the full sample with the semi-analytical simulation of cosmological galaxy formation by \citet{croton06}. They examined two different modes of black hole accretion, namely quasar and radio mode. They found that powerful outflows which occur in radio mode accretion can efficiently quench star formation in massive systems, thus offering a solution for the cooling flow problem. One of the predictions of their model is the BH accretion rate density evolution, estimated using the Bondi-Hoyle accretion model. An update to this work has been presented in \citet{croton16} where the authors introduced the so called `radio mode efficiency' factor to modulate the strength of the black hole accretion. Using the accretion rate formula (see Eq. 16 in \citealt{croton16}), and adopting the suggested radio mode efficiency parameter value of 0.08, the black hole luminosity can be calculated. Under the assumption that most of the accretion energy is ejected in the kinetic form, we can compare the \citeauthor{croton06} black hole luminosity to our estimate of KLD for the full sample which contains the contribution from all AGN that show evidence of AGN-related radio emission. At redshifts higher than $\mathrm{z\sim1.5}$ the full sample KLD curve shows a decline similar to that expected from semi-analytic simulation by \citeauthor{croton16} for the radio mode AGN. The observed decline towards the local value at lower redshift instead is not in agreement with the predicted radio mode curve. As discussed, the normalization of KLD curves is affected by the uncertainty on the $\mathrm{f_{W}}$ factor. The normalization of the peak of the KLD curve of the full sample would agree with the \citeauthor{croton16} expectations for $\mathrm{f_{W} = 4}$.

From Fig. \ref{fig:mech_power} it is clear that the two simulations compared disagree in their prediction of the KLD both in normalization and overall shape. This stresses the need for more in-depth studies of the conversion between monochromatic and kinetic luminosity.

In summary, we find that our data do not completely agree with any of the predictions from the available models. However, given the significant uncertainties present both in the data and in the models, it is interesting to note that each of the features shown by the data (i.e. approximately constant maximum of KLD in the redshift range $\mathrm{1 < z < 2.5}$, with significant decreases at both lower and higher redshift), are predicted by at least one of the models. This suggest that, using somewhat different assumptions, it should be possible to generate new models of galaxy evolution which could better reproduce our data.

\section{Summary and conclusions} \label{sec:sac}

Here we present an analysis of a sample of $1,604$ X-ray and mid-IR selected AGN with moderate-to-high radiative luminosities (HLAGN) from the VLA-COSMOS 3 GHz Large Project, selected to trace analogs of local high-excitation radio galaxies (HERGs) out to $\mathrm{z\sim6}$. Our findings are summarized as follows.

\begin{description}
  \item[$\bullet$]  To study the origin of the radio emission in our HLAGN sample, we develop the method of statistical decomposition of the observed 3 GHz radio luminosities into contributions arising from star-forming processes and AGN activity. In agreement with other studies of X-ray and MIR selected AGN samples, we find that the majority, $(68.0\pm1.5)\%$, of HLAGN have radio emission dominated by star-forming processes ($\mathrm{0 \leq f_{AGN}\leq 0.5}$), while $(32.0\pm1.5)\%$ of them are dominated by the AGN-related emission in the radio ($0.5<\mathrm{f_{AGN}\leq1.0}$).

\item[$\bullet$]  For all sources with $\mathrm{f_{AGN}>0}$ we calculate the AGN luminosity by scaling the total 1.4 GHz luminosity (converted from the observed 3 GHz radio luminosity) by the derived AGN fraction. Using a complete subsample ($\mathrm{S_{3\ GHz, AGN}\geq S_{lim, 3\ GHz}}$) of HLAGN we construct the 1.4 GHz radio AGN luminosity function of HLAGN out to $\mathrm{z\sim6}$. By assuming pure luminosity and pure density evolution models we find $\mathrm{L^* (z) \propto (1+z)^{ (3.97\pm 0.15) + (-0.92\pm 0.06) z}}$ and $\mathrm{\Phi^* (z) \propto (1+z)^{(2.64\pm 0.10) + (-0.61\pm 0.04) z}}$, respectively.

\item[$\bullet$]  We further study the evolution of the number and luminosity densities of the HLAGN population. We find an increasing trend from high redshifts up to a flattening at redshifts $\mathrm{1<z<2.5}$, followed by a significant decrease down to the value in the local Universe.

\item[$\bullet$]  We estimate the kinetic luminosity density and find an approximately constant maximum in the redshift range $\mathrm{1 < z < 2.5}$. By comparing this behavior to expectations from semi-analytic models of galaxy formation and evolution, we find that such behavior is not entirely expected by any of the models tested here. However, given the significant uncertainties present both in the data and in the models, the similarities found show that it should be possible, by modifying assumptions of these models, to better reproduce our data.

\end{description}

\

\

\textit{Acknowledgements.} The authors are grateful to the referee for comments which helped to improve the content of this article. LC is grateful to A. Griffin for useful suggestions. This research was founded by the European Union's Seventh Frame-work program under grant agreement 337595 (ERC Starting Grant, 'CoSMass'). JD acknowledges financial assistance from the South African SKA Project (SKA SA; www.ska.ac.za). EV acknowledges funding from the DFG grant BE 1837/13-1 and support of the Collaborative Research Center 956, subproject A1 and C4, funded by the Deutsche Forschungsgemeinschaft (DFG).

\bibliographystyle{apj}
\bibliography{bibliography}

\begin{thebibliography}{}
\expandafter\ifx\csname natexlab\endcsname\relax\def\natexlab#1{#1}\fi

\bibitem[{{Becker} {et~al.}(1994){Becker}, {White}, \& {Helfand}}]{becker94}
{Becker}, R.~H., {White}, R.~L., \& {Helfand}, D.~J. 1994, in Astronomical
  Society of the Pacific Conference Series, Vol.~61, Astronomical Data Analysis
  Software and Systems III, ed. D.~R. {Crabtree}, R.~J. {Hanisch}, \&
  J.~{Barnes}, 165

\bibitem[{{Bell}(2003)}]{bell03}
{Bell}, E.~F. 2003, \apj, 586, 794

\bibitem[{{Best} \& {Heckman}(2012)}]{bh12}
{Best}, P.~N., \& {Heckman}, T.~M. 2012, \mnras, 421, 1569

\bibitem[{{Best} {et~al.}(2014){Best}, {Ker}, {Simpson}, {Rigby}, \&
  {Sabater}}]{best14}
{Best}, P.~N., {Ker}, L.~M., {Simpson}, C., {Rigby}, E.~E., \& {Sabater}, J.
  2014, \mnras, 445, 955

\bibitem[{{B{\^i}rzan} {et~al.}(2004){B{\^i}rzan}, {Rafferty}, {McNamara},
  {Wise}, \& {Nulsen}}]{birzan04}
{B{\^i}rzan}, L., {Rafferty}, D.~A., {McNamara}, B.~R., {Wise}, M.~W., \&
  {Nulsen}, P.~E.~J. 2004, \apj, 607, 800

\bibitem[{{Blandford} \& {Znajek}(1977)}]{bz77}
{Blandford}, R.~D., \& {Znajek}, R.~L. 1977, \mnras, 179, 433

\bibitem[{{Bonzini} {et~al.}(2013){Bonzini}, {Padovani}, {Mainieri},
  {Kellermann}, {Miller}, {Rosati}, {Tozzi}, \& {Vattakunnel}}]{bonzini13}
{Bonzini}, M., {Padovani}, P., {Mainieri}, V., {et~al.} 2013, \mnras, 436, 3759

\bibitem[{{Bower} {et~al.}(2006){Bower}, {Benson}, {Malbon}, {Helly}, {Frenk},
  {Baugh}, {Cole}, \& {Lacey}}]{bower06}
{Bower}, R.~G., {Benson}, A.~J., {Malbon}, R., {et~al.} 2006, \mnras, 370, 645

\bibitem[{{Brinchmann} {et~al.}(2004){Brinchmann}, {Charlot}, {White},
  {Tremonti}, {Kauffmann}, {Heckman}, \& {Brinkmann}}]{brinchmann04}
{Brinchmann}, J., {Charlot}, S., {White}, S.~D.~M., {et~al.} 2004, \mnras, 351,
  1151

\bibitem[{{Capak} {et~al.}(2007){Capak}, {Aussel}, {Ajiki}, {McCracken},
  {Mobasher}, {Scoville}, {Shopbell}, {Taniguchi}, {Thompson}, {Tribiano},
  {Sasaki}, {Blain}, {Brusa}, {Carilli}, {Comastri}, {Carollo}, {Cassata},
  {Colbert}, {Ellis}, {Elvis}, {Giavalisco}, {Green}, {Guzzo}, {Hasinger},
  {Ilbert}, {Impey}, {Jahnke}, {Kartaltepe}, {Kneib}, {Koda}, {Koekemoer},
  {Komiyama}, {Leauthaud}, {Le Fevre}, {Lilly}, {Liu}, {Massey}, {Miyazaki},
  {Murayama}, {Nagao}, {Peacock}, {Pickles}, {Porciani}, {Renzini}, {Rhodes},
  {Rich}, {Salvato}, {Sanders}, {Scarlata}, {Schiminovich}, {Schinnerer},
  {Scodeggio}, {Sheth}, {Shioya}, {Tasca}, {Taylor}, {Yan}, \&
  {Zamorani}}]{capak07}
{Capak}, P., {Aussel}, H., {Ajiki}, M., {et~al.} 2007, \apjs, 172, 99

\bibitem[{{Cavagnolo} {et~al.}(2010){Cavagnolo}, {McNamara}, {Nulsen},
  {Carilli}, {Jones}, \& {B{\^i}rzan}}]{cavagnolo10}
{Cavagnolo}, K.~W., {McNamara}, B.~R., {Nulsen}, P.~E.~J., {et~al.} 2010, \apj,
  720, 1066

\bibitem[{{Chabrier}(2003)}]{chabrier}
{Chabrier}, G. 2003, \pasp, 115, 763

\bibitem[{{Ching} {et~al.}(2017){Ching}, {Sadler}, {Croom}, {Johnston},
  {Pracy}, {Couch}, {Hopkins}, {Jurek}, \& {Pimbblet}}]{ching17}
{Ching}, J.~H.~Y., {Sadler}, E.~M., {Croom}, S.~M., {et~al.} 2017, \mnras, 464,
  1306

\bibitem[{{Civano} {et~al.}(2016){Civano}, {Marchesi}, {Comastri}, {Urry},
  {Elvis}, {Cappelluti}, {Puccetti}, {Brusa}, {Zamorani}, {Hasinger},
  {Aldcroft}, {Alexander}, {Allevato}, {Brunner}, {Capak}, {Finoguenov},
  {Fiore}, {Fruscione}, {Gilli}, {Glotfelty}, {Griffiths}, {Hao}, {Harrison},
  {Jahnke}, {Kartaltepe}, {Karim}, {LaMassa}, {Lanzuisi}, {Miyaji}, {Ranalli},
  {Salvato}, {Sargent}, {Scoville}, {Schawinski}, {Schinnerer}, {Silverman},
  {Smolcic}, {Stern}, {Toft}, {Trakhtenbrot}, {Treister}, \&
  {Vignali}}]{civano16}
{Civano}, F., {Marchesi}, S., {Comastri}, A., {et~al.} 2016, \apj, 819, 62

\bibitem[{{Condon}(1992)}]{condon92}
{Condon}, J.~J. 1992, \araa, 30, 575

\bibitem[{{Condon} {et~al.}(1998){Condon}, {Cotton}, {Greisen}, {Yin},
  {Perley}, {Taylor}, \& {Broderick}}]{condon98}
{Condon}, J.~J., {Cotton}, W.~D., {Greisen}, E.~W., {et~al.} 1998, \aj, 115,
  1693

\bibitem[{{Croton} {et~al.}(2006){Croton}, {Springel}, {White}, {De Lucia},
  {Frenk}, {Gao}, {Jenkins}, {Kauffmann}, {Navarro}, \& {Yoshida}}]{croton06}
{Croton}, D.~J., {Springel}, V., {White}, S.~D.~M., {et~al.} 2006, \mnras, 365,
  11

\bibitem[{{Croton} {et~al.}(2016){Croton}, {Stevens}, {Tonini}, {Garel},
  {Bernyk}, {Bibiano}, {Hodkinson}, {Mutch}, {Poole}, \& {Shattow}}]{croton16}
{Croton}, D.~J., {Stevens}, A.~R.~H., {Tonini}, C., {et~al.} 2016, \apjs, 222,
  22

\bibitem[{{Delhaize} {et~al.}(2017){Delhaize}, {Smol{\v c}i{\'c}},
  {Delvecchio}, {Novak}, {Sargent}, {Baran}, {Magnelli}, {Zamorani},
  {Schinnerer}, {Murphy}, {Aravena}, {Berta}, {Bondi}, {Capak}, {Carilli},
  {Ciliegi}, {Civano}, {Ilbert}, {Karim}, {Laigle}, {Le F{\`e}vre}, {Marchesi},
  {McCracken}, {Salvato}, {Seymour}, \& {Tasca}}]{delhaize17}
{Delhaize}, J., {Smol{\v c}i{\'c}}, V., {Delvecchio}, I., {et~al.} 2017, \aap,
  602, A4

\bibitem[{{Delvecchio} {et~al.}(2017){Delvecchio}, {Smol{\v c}i{\'c}},
  {Zamorani}, {Lagos}, {Berta}, {Delhaize}, {Baran}, {Alexander}, {Rosario},
  {Gonzalez-Perez}, {Ilbert}, {Lacey}, {Le F{\`e}vre}, {Miettinen}, {Aravena},
  {Bondi}, {Carilli}, {Ciliegi}, {Mooley}, {Novak}, {Schinnerer}, {Capak},
  {Civano}, {Fanidakis}, {Herrera Ruiz}, {Karim}, {Laigle}, {Marchesi},
  {McCracken}, {Middleberg}, {Salvato}, \& {Tasca}}]{delvecchio17}
{Delvecchio}, I., {Smol{\v c}i{\'c}}, V., {Zamorani}, G., {et~al.} 2017, \aap,
  602, A3

\bibitem[{{Donley} {et~al.}(2012){Donley}, {Koekemoer}, {Brusa}, {Capak},
  {Cardamone}, {Civano}, {Ilbert}, {Impey}, {Kartaltepe}, {Miyaji}, {Salvato},
  {Sanders}, {Trump}, \& {Zamorani}}]{donley12}
{Donley}, J.~L., {Koekemoer}, A.~M., {Brusa}, M., {et~al.} 2012, \apj, 748, 142

\bibitem[{{Dunlop} {et~al.}(2003){Dunlop}, {McLure}, {Kukula}, {Baum}, {O'Dea},
  \& {Hughes}}]{dunlop03}
{Dunlop}, J.~S., {McLure}, R.~J., {Kukula}, M.~J., {et~al.} 2003, \mnras, 340,
  1095

\bibitem[{{Fabian}(2012)}]{fabian12}
{Fabian}, A.~C. 2012, \araa, 50, 455

\bibitem[{{Falcke} \& {Biermann}(1999)}]{fb99}
{Falcke}, H., \& {Biermann}, P.~L. 1999, \aap, 342, 49

\bibitem[{{Fanidakis} {et~al.}(2011){Fanidakis}, {Baugh}, {Benson}, {Bower},
  {Cole}, {Done}, \& {Frenk}}]{fanidakis11}
{Fanidakis}, N., {Baugh}, C.~M., {Benson}, A.~J., {et~al.} 2011, \mnras, 410,
  53

\bibitem[{{Fanidakis} {et~al.}(2012){Fanidakis}, {Baugh}, {Benson}, {Bower},
  {Cole}, {Done}, {Frenk}, {Hickox}, {Lacey}, \& {Del P.~Lagos}}]{fanidakis12}
---. 2012, \mnras, 419, 2797

\bibitem[{{Ferrarese} \& {Merritt}(2000)}]{fm00}
{Ferrarese}, L., \& {Merritt}, D. 2000, \apjl, 539, L9

\bibitem[{{Filho} {et~al.}(2006){Filho}, {Barthel}, \& {Ho}}]{filho06}
{Filho}, M.~E., {Barthel}, P.~D., \& {Ho}, L.~C. 2006, \aap, 451, 71

\bibitem[{{Freedman} \& {Diaconis}(1981)}]{fd81a}
{Freedman}, D., \& {Diaconis}, P. 1981, Probability Theory and Related Fields,
  57, 453

\bibitem[{{Gehrels}(1986)}]{gehrels86}
{Gehrels}, N. 1986, \apj, 303, 336

\bibitem[{{Gendre} {et~al.}(2013){Gendre}, {Best}, {Wall}, \& {Ker}}]{gendre13}
{Gendre}, M.~A., {Best}, P.~N., {Wall}, J.~V., \& {Ker}, L.~M. 2013, \mnras,
  430, 3086

\bibitem[{{Godfrey} \& {Shabala}(2016)}]{gs16}
{Godfrey}, L.~E.~H., \& {Shabala}, S.~S. 2016, \mnras, 456, 1172

\bibitem[{{Goulding} {et~al.}(2014){Goulding}, {Forman}, {Hickox}, {Jones},
  {Murray}, {Paggi}, {Ashby}, {Coil}, {Cooper}, {Huang}, {Kraft}, {Newman},
  {Weiner}, \& {Willner}}]{goulding14}
{Goulding}, A.~D., {Forman}, W.~R., {Hickox}, R.~C., {et~al.} 2014, \apj, 783,
  40

\bibitem[{{Guzzo} {et~al.}(2007){Guzzo}, {Cassata}, {Finoguenov}, {Massey},
  {Scoville}, {Capak}, {Ellis}, {Mobasher}, {Taniguchi}, {Thompson}, {Ajiki},
  {Aussel}, {B{\"o}hringer}, {Brusa}, {Calzetti}, {Comastri}, {Franceschini},
  {Hasinger}, {Kasliwal}, {Kitzbichler}, {Kneib}, {Koekemoer}, {Leauthaud},
  {McCracken}, {Murayama}, {Nagao}, {Rhodes}, {Sanders}, {Sasaki}, {Shioya},
  {Tasca}, \& {Taylor}}]{guzzo07}
{Guzzo}, L., {Cassata}, P., {Finoguenov}, A., {et~al.} 2007, \apjs, 172, 254

\bibitem[{{Hardcastle} \& {Krause}(2014)}]{hardcastle14}
{Hardcastle}, M.~J., \& {Krause}, M.~G.~H. 2014, \mnras, 443, 1482

\bibitem[{{Heckman} \& {Best}(2014)}]{hb14}
{Heckman}, T.~M., \& {Best}, P.~N. 2014, \araa, 52, 589

\bibitem[{{Herrera Ruiz} {et~al.}(2017){Herrera Ruiz}, {Middelberg}, {Deller},
  {Norris}, {Best}, {Brisken}, {Schinnerer}, {Smol{\v c}i{\'c}}, {Delvecchio},
  {Momjian}, {Bomans}, {Scoville}, \& {Carilli}}]{herreraruiz17}
{Herrera Ruiz}, N., {Middelberg}, E., {Deller}, A., {et~al.} 2017, \aap, 607,
  A132

\bibitem[{{Hickox} {et~al.}(2009){Hickox}, {Jones}, {Forman}, {Murray},
  {Kochanek}, {Eisenstein}, {Jannuzi}, {Dey}, {Brown}, {Stern}, {Eisenhardt},
  {Gorjian}, {Brodwin}, {Narayan}, {Cool}, {Kenter}, {Caldwell}, \&
  {Anderson}}]{hickox09}
{Hickox}, R.~C., {Jones}, C., {Forman}, W.~R., {et~al.} 2009, \apj, 696, 891

\bibitem[{{Ho} {et~al.}(1997){Ho}, {Filippenko}, \& {Sargent}}]{palomar97}
{Ho}, L.~C., {Filippenko}, A.~V., \& {Sargent}, W.~L.~W. 1997, \apjs, 112, 315

\bibitem[{{Iovino} {et~al.}(2016){Iovino}, {Petropoulou}, {Scodeggio},
  {Bolzonella}, {Zamorani}, {Bardelli}, {Cucciati}, {Pozzetti}, {Tasca},
  {Vergani}, {Zucca}, {Finoguenov}, {Ilbert}, {Tanaka}, {Salvato}, {Kova{\v
  c}}, \& {Cassata}}]{iovino16}
{Iovino}, A., {Petropoulou}, V., {Scodeggio}, M., {et~al.} 2016, \aap, 592, A78

\bibitem[{{Kapinska} {et~al.}(2015){Kapinska}, {Hardcastle}, {Jackson}, {An},
  {Baan}, \& {Jarvis}}]{kapinska14}
{Kapinska}, A.~D., {Hardcastle}, M., {Jackson}, C., {et~al.} 2015, Advancing
  Astrophysics with the Square Kilometre Array (AASKA14), 173

\bibitem[{{Kennicutt}(1998)}]{kennicutt98}
{Kennicutt}, Jr., R.~C. 1998, \apj, 498, 541

\bibitem[{{King}(2008)}]{king08}
{King}, A. 2008, \memsai, 79, 1066

\bibitem[{{Lacey} {et~al.}(2016){Lacey}, {Baugh}, {Frenk}, {Benson}, {Bower},
  {Cole}, {Gonzalez-Perez}, {Helly}, {Lagos}, \& {Mitchell}}]{lacey16}
{Lacey}, C.~G., {Baugh}, C.~M., {Frenk}, C.~S., {et~al.} 2016, \mnras, 462,
  3854

\bibitem[{{Laigle} {et~al.}(2016){Laigle}, {McCracken}, {Ilbert}, {Hsieh},
  {Davidzon}, {Capak}, {Hasinger}, {Silverman}, {Pichon}, {Coupon}, {Aussel},
  {Le Borgne}, {Caputi}, {Cassata}, {Chang}, {Civano}, {Dunlop}, {Fynbo},
  {Kartaltepe}, {Koekemoer}, {Le F{\`e}vre}, {Le Floc'h}, {Leauthaud}, {Lilly},
  {Lin}, {Marchesi}, {Milvang-Jensen}, {Salvato}, {Sanders}, {Scoville},
  {Smolcic}, {Stockmann}, {Taniguchi}, {Tasca}, {Toft}, {Vaccari}, \&
  {Zabl}}]{laigle16}
{Laigle}, C., {McCracken}, H.~J., {Ilbert}, O., {et~al.} 2016, \apjs, 224, 24

\bibitem[{{Magnelli} {et~al.}(2015){Magnelli}, {Ivison}, {Lutz}, {Valtchanov},
  {Farrah}, {Berta}, {Bertoldi}, {Bock}, {Cooray}, {Ibar}, {Karim}, {Le
  Floc'h}, {Nordon}, {Oliver}, {Page}, {Popesso}, {Pozzi}, {Rigopoulou},
  {Riguccini}, {Rodighiero}, {Rosario}, {Roseboom}, {Wang}, \&
  {Wuyts}}]{magnelli15}
{Magnelli}, B., {Ivison}, R.~J., {Lutz}, D., {et~al.} 2015, \aap, 573, A45

\bibitem[{{Magorrian} {et~al.}(1998){Magorrian}, {Tremaine}, {Richstone},
  {Bender}, {Bower}, {Dressler}, {Faber}, {Gebhardt}, {Green}, {Grillmair},
  {Kormendy}, \& {Lauer}}]{magorrian98}
{Magorrian}, J., {Tremaine}, S., {Richstone}, D., {et~al.} 1998, \aj, 115, 2285

\bibitem[{{Marchesi} {et~al.}(2016){Marchesi}, {Civano}, {Elvis}, {Salvato},
  {Brusa}, {Comastri}, {Gilli}, {Hasinger}, {Lanzuisi}, {Miyaji}, {Treister},
  {Urry}, {Vignali}, {Zamorani}, {Allevato}, {Cappelluti}, {Cardamone},
  {Finoguenov}, {Griffiths}, {Karim}, {Laigle}, {LaMassa}, {Jahnke}, {Ranalli},
  {Schawinski}, {Schinnerer}, {Silverman}, {Smolcic}, {Suh}, \&
  {Trakhtenbrot}}]{marchesi16}
{Marchesi}, S., {Civano}, F., {Elvis}, M., {et~al.} 2016, \apj, 817, 34

\bibitem[{{Marshall}(1985)}]{marshall85}
{Marshall}, H.~L. 1985, \apj, 299, 109

\bibitem[{{Mauch} \& {Sadler}(2007)}]{ms07}
{Mauch}, T., \& {Sadler}, E.~M. 2007, \mnras, 375, 931

\bibitem[{{Merloni} \& {Heinz}(2007)}]{mh07}
{Merloni}, A., \& {Heinz}, S. 2007, \mnras, 381, 589

\bibitem[{{Merloni} \& {Heinz}(2008)}]{mh08}
---. 2008, \mnras, 388, 1011

\bibitem[{{Miller} {et~al.}(1993){Miller}, {Rawlings}, \&
  {Saunders}}]{miller93}
{Miller}, P., {Rawlings}, S., \& {Saunders}, R. 1993, \mnras, 263, 425

\bibitem[{{Narayan} \& {Yi}(1994)}]{ny94}
{Narayan}, R., \& {Yi}, I. 1994, \apjl, 428, L13

\bibitem[{{Narayan} \& {Yi}(1995)}]{ny95}
---. 1995, \apj, 444, 231

\bibitem[{{Novak} {et~al.}(2018){Novak}, {Smolcic}, {Schinnerer}, {Zamorani},
  {Delvecchio}, {Bondi}, \& {Delhaize}}]{novak18}
{Novak}, M., {Smolcic}, V., {Schinnerer}, E., {et~al.} 2018, ArXiv e-prints,
  arXiv:1803.01569

\bibitem[{{Novak} {et~al.}(2017){Novak}, {Smol{\v c}i{\'c}}, {Delhaize},
  {Delvecchio}, {Zamorani}, {Baran}, {Bondi}, {Capak}, {Carilli}, {Ciliegi},
  {Civano}, {Ilbert}, {Karim}, {Laigle}, {Le F{\`e}vre}, {Marchesi},
  {McCracken}, {Miettinen}, {Salvato}, {Sargent}, {Schinnerer}, \&
  {Tasca}}]{novak17}
{Novak}, M., {Smol{\v c}i{\'c}}, V., {Delhaize}, J., {et~al.} 2017, \aap, 602,
  A5

\bibitem[{{O'Sullivan} {et~al.}(2011){O'Sullivan}, {Giacintucci}, {David},
  {Gitti}, {Vrtilek}, {Raychaudhury}, \& {Ponman}}]{o'sullivan11}
{O'Sullivan}, E., {Giacintucci}, S., {David}, L.~P., {et~al.} 2011, \apj, 735,
  11

\bibitem[{{Padovani} {et~al.}(2015){Padovani}, {Bonzini}, {Kellermann},
  {Miller}, {Mainieri}, \& {Tozzi}}]{padovani15}
{Padovani}, P., {Bonzini}, M., {Kellermann}, K.~I., {et~al.} 2015, \mnras, 452,
  1263

\bibitem[{{Pracy} {et~al.}(2016){Pracy}, {Ching}, {Sadler}, {Croom}, {Baldry},
  {Bland-Hawthorn}, {Brough}, {Brown}, {Couch}, {Davis}, {Drinkwater},
  {Hopkins}, {Jarvis}, {Jelliffe}, {Jurek}, {Loveday}, {Pimbblet}, {Prescott},
  {Wisnioski}, \& {Woods}}]{pracy16}
{Pracy}, M.~B., {Ching}, J.~H.~Y., {Sadler}, E.~M., {et~al.} 2016, \mnras, 460,
  2

\bibitem[{{Rigby} {et~al.}(2015){Rigby}, {Argyle}, {Best}, {Rosario}, \&
  {R{\"o}ttgering}}]{rigby15}
{Rigby}, E.~E., {Argyle}, J., {Best}, P.~N., {Rosario}, D., \&
  {R{\"o}ttgering}, H.~J.~A. 2015, \aap, 581, A96

\bibitem[{{Salim} {et~al.}(2007){Salim}, {Rich}, {Charlot}, {Brinchmann},
  {Johnson}, {Schiminovich}, {Seibert}, {Mallery}, {Heckman}, {Forster},
  {Friedman}, {Martin}, {Morrissey}, {Neff}, {Small}, {Wyder}, {Bianchi},
  {Donas}, {Lee}, {Madore}, {Milliard}, {Szalay}, {Welsh}, \& {Yi}}]{salim07}
{Salim}, S., {Rich}, R.~M., {Charlot}, S., {et~al.} 2007, \apjs, 173, 267

\bibitem[{{Sargent} {et~al.}(2010){Sargent}, {Schinnerer}, {Murphy}, {Aussel},
  {Le Floc'h}, {Frayer}, {Mart{\'{\i}}nez-Sansigre}, {Oesch}, {Salvato},
  {Smol{\v c}i{\'c}}, {Zamorani}, {Brusa}, {Cappelluti}, {Carilli}, {Carollo},
  {Ilbert}, {Kartaltepe}, {Koekemoer}, {Lilly}, {Sanders}, \&
  {Scoville}}]{sargent10}
{Sargent}, M.~T., {Schinnerer}, E., {Murphy}, E., {et~al.} 2010, \apjs, 186,
  341

\bibitem[{{Schinnerer} {et~al.}(2007){Schinnerer}, {Smol{\v c}i{\'c}},
  {Carilli}, {Bondi}, {Ciliegi}, {Jahnke}, {Scoville}, {Aussel}, {Bertoldi},
  {Blain}, {Impey}, {Koekemoer}, {Le Fevre}, \& {Urry}}]{schinnerer07}
{Schinnerer}, E., {Smol{\v c}i{\'c}}, V., {Carilli}, C.~L., {et~al.} 2007,
  \apjs, 172, 46

\bibitem[{{Schinnerer} {et~al.}(2010){Schinnerer}, {Sargent}, {Bondi}, {Smol{\v
  c}i{\'c}}, {Datta}, {Carilli}, {Bertoldi}, {Blain}, {Ciliegi}, {Koekemoer},
  \& {Scoville}}]{schinnerer10}
{Schinnerer}, E., {Sargent}, M.~T., {Bondi}, M., {et~al.} 2010, \apjs, 188, 384

\bibitem[{{Schmidt}(1968)}]{schmidt68}
{Schmidt}, M. 1968, \apj, 151, 393

\bibitem[{{Shakura} \& {Sunyaev}(1973)}]{ss73}
{Shakura}, N.~I., \& {Sunyaev}, R.~A. 1973, \aap, 24, 337

\bibitem[{{Smolcic}(2016)}]{smolcic16}
{Smolcic}, V. 2016, ArXiv e-prints, arXiv:1603.05687

\bibitem[{{Smol{\v c}i{\'c}} {et~al.}(2009){Smol{\v c}i{\'c}}, {Zamorani},
  {Schinnerer}, {Bardelli}, {Bondi}, {B{\^i}rzan}, {Carilli}, {Ciliegi},
  {Elvis}, {Impey}, {Koekemoer}, {Merloni}, {Paglione}, {Salvato}, {Scodeggio},
  {Scoville}, \& {Trump}}]{smolcic09}
{Smol{\v c}i{\'c}}, V., {Zamorani}, G., {Schinnerer}, E., {et~al.} 2009, \apj,
  696, 24

\bibitem[{{Smol{\v c}i{\'c}} {et~al.}(2017{\natexlab{a}}){Smol{\v c}i{\'c}},
  {Novak}, {Bondi}, {Ciliegi}, {Mooley}, {Schinnerer}, {Zamorani}, {Navarrete},
  {Bourke}, {Karim}, {Vardoulaki}, {Leslie}, {Delhaize}, {Carilli}, {Myers},
  {Baran}, {Delvecchio}, {Miettinen}, {Banfield}, {Balokovi{\'c}}, {Bertoldi},
  {Capak}, {Frail}, {Hallinan}, {Hao}, {Herrera Ruiz}, {Horesh}, {Ilbert},
  {Intema}, {Jeli{\'c}}, {Kl{\"o}ckner}, {Krpan}, {Kulkarni}, {McCracken},
  {Laigle}, {Middleberg}, {Murphy}, {Sargent}, {Scoville}, \&
  {Sheth}}]{smolcic17a}
{Smol{\v c}i{\'c}}, V., {Novak}, M., {Bondi}, M., {et~al.} 2017{\natexlab{a}},
  \aap, 602, A1

\bibitem[{{Smol{\v c}i{\'c}} {et~al.}(2017{\natexlab{b}}){Smol{\v c}i{\'c}},
  {Novak}, {Delvecchio}, {Ceraj}, {Bondi}, {Delhaize}, {Marchesi}, {Murphy},
  {Schinnerer}, {Vardoulaki}, \& {Zamorani}}]{smolcic17c}
{Smol{\v c}i{\'c}}, V., {Novak}, M., {Delvecchio}, I., {et~al.}
  2017{\natexlab{b}}, \aap, 602, A6

\bibitem[{{Smol{\v c}i{\'c}} {et~al.}(2017{\natexlab{c}}){Smol{\v c}i{\'c}},
  {Delvecchio}, {Zamorani}, {Baran}, {Novak}, {Delhaize}, {Schinnerer},
  {Berta}, {Bondi}, {Ciliegi}, {Capak}, {Civano}, {Karim}, {Le Fevre},
  {Ilbert}, {Laigle}, {Marchesi}, {McCracken}, {Tasca}, {Salvato}, \&
  {Vardoulaki}}]{smolcic17b}
{Smol{\v c}i{\'c}}, V., {Delvecchio}, I., {Zamorani}, G., {et~al.}
  2017{\natexlab{c}}, \aap, 602, A2

\bibitem[{{Steinhardt} {et~al.}(2014){Steinhardt}, {Speagle}, {Capak},
  {Silverman}, {Carollo}, {Dunlop}, {Hashimoto}, {Hsieh}, {Ilbert}, {Le Fevre},
  {Le Floc'h}, {Lee}, {Lin}, {Lin}, {Masters}, {McCracken}, {Nagao}, {Petric},
  {Salvato}, {Sanders}, {Scoville}, {Sheth}, {Strauss}, \&
  {Taniguchi}}]{steinhardt14}
{Steinhardt}, C.~L., {Speagle}, J.~S., {Capak}, P., {et~al.} 2014, \apjl, 791,
  L25

\bibitem[{{Whitaker} {et~al.}(2012){Whitaker}, {van Dokkum}, {Brammer}, \&
  {Franx}}]{whitaker12}
{Whitaker}, K.~E., {van Dokkum}, P.~G., {Brammer}, G., \& {Franx}, M. 2012,
  \apjl, 754, L29

\bibitem[{{Willott} {et~al.}(1999){Willott}, {Rawlings}, {Blundell}, \&
  {Lacy}}]{willott99}
{Willott}, C.~J., {Rawlings}, S., {Blundell}, K.~M., \& {Lacy}, M. 1999,
  \mnras, 309, 1017

\bibitem[{{York} {et~al.}(2000){York}, {Adelman}, {Anderson}, {Anderson},
  {Annis}, {Bahcall}, {Bakken}, {Barkhouser}, {Bastian}, {Berman}, {Boroski},
  {Bracker}, {Briegel}, {Briggs}, {Brinkmann}, {Brunner}, {Burles}, {Carey},
  {Carr}, {Castander}, {Chen}, {Colestock}, {Connolly}, {Crocker}, {Csabai},
  {Czarapata}, {Davis}, {Doi}, {Dombeck}, {Eisenstein}, {Ellman}, {Elms},
  {Evans}, {Fan}, {Federwitz}, {Fiscelli}, {Friedman}, {Frieman}, {Fukugita},
  {Gillespie}, {Gunn}, {Gurbani}, {de Haas}, {Haldeman}, {Harris}, {Hayes},
  {Heckman}, {Hennessy}, {Hindsley}, {Holm}, {Holmgren}, {Huang}, {Hull},
  {Husby}, {Ichikawa}, {Ichikawa}, {Ivezi{\'c}}, {Kent}, {Kim}, {Kinney},
  {Klaene}, {Kleinman}, {Kleinman}, {Knapp}, {Korienek}, {Kron}, {Kunszt},
  {Lamb}, {Lee}, {Leger}, {Limmongkol}, {Lindenmeyer}, {Long}, {Loomis},
  {Loveday}, {Lucinio}, {Lupton}, {MacKinnon}, {Mannery}, {Mantsch}, {Margon},
  {McGehee}, {McKay}, {Meiksin}, {Merelli}, {Monet}, {Munn}, {Narayanan},
  {Nash}, {Neilsen}, {Neswold}, {Newberg}, {Nichol}, {Nicinski}, {Nonino},
  {Okada}, {Okamura}, {Ostriker}, {Owen}, {Pauls}, {Peoples}, {Peterson},
  {Petravick}, {Pier}, {Pope}, {Pordes}, {Prosapio}, {Rechenmacher}, {Quinn},
  {Richards}, {Richmond}, {Rivetta}, {Rockosi}, {Ruthmansdorfer}, {Sandford},
  {Schlegel}, {Schneider}, {Sekiguchi}, {Sergey}, {Shimasaku}, {Siegmund},
  {Smee}, {Smith}, {Snedden}, {Stone}, {Stoughton}, {Strauss}, {Stubbs},
  {SubbaRao}, {Szalay}, {Szapudi}, {Szokoly}, {Thakar}, {Tremonti}, {Tucker},
  {Uomoto}, {Vanden Berk}, {Vogeley}, {Waddell}, {Wang}, {Watanabe},
  {Weinberg}, {Yanny}, {Yasuda}, \& {SDSS Collaboration}}]{york00}
{York}, D.~G., {Adelman}, J., {Anderson}, Jr., J.~E., {et~al.} 2000, \aj, 120,
  1579

\bibitem[{{Yun} {et~al.}(2001){Yun}, {Reddy}, \& {Condon}}]{yun01}
{Yun}, M.~S., {Reddy}, N.~A., \& {Condon}, J.~J. 2001, \apj, 554, 803

\end{thebibliography}

\appendix
\clearpage
\section{Analysis of the full radio sample in the COSMOS field}
\label{sec:comp_c15}

In Sec. \ref{sec:decomposition} we introduced the method of statistically decomposing the total radio luminosity of HLAGN into the components arising from star-forming processes and from AGN activity. In this section we describe the calibration of the star formation-related infrared-to-1.4 GHz luminosity ratio, $\mathrm{q_{SF}}$, on the full sample of the radio detected sources from the VLA-COSMOS 3 GHz Large Project with COSMOS2015 counterparts. We further use the derived calibration to decompose the total radio luminosities of the full sample and constrain the evolution of its radio AGN luminosity function.

\subsection{Calibration of the infrared-to-1.4 GHz luminosity ratio of the host galaxy} \label{sec:calibration}

The full sample of the radio detected sources from the VLA-COSMOS 3 GHz Large Project with COSMOS2015 counterparts contains 7,729 sources, classified as HLAGN, MLAGN or SFGs (see Sec. \ref{sec:hlagn}). We further refer to this sample as the full sample. Since we aim at isolating the AGN-related radio emission, here we use the decomposition technique to derive AGN fractions for all sources which display radio excess, assuming this excess arises from AGN activity.

\begin{figure}[h]
 \includegraphics[width=\linewidth]{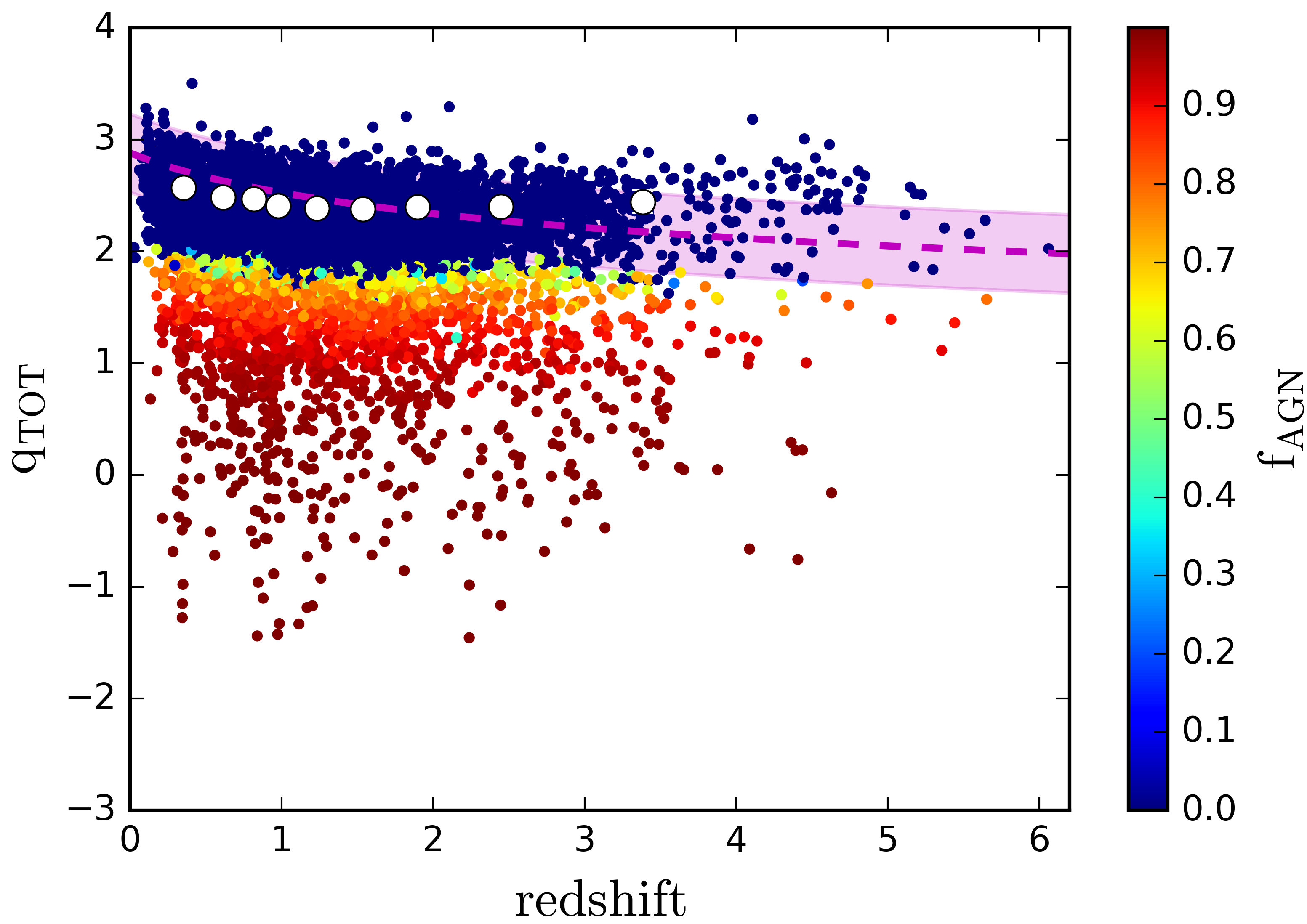}
  \caption{Infrared-to-1.4 GHz luminosity ratio vs. redshift, color-coded by AGN fractions. The magenta dashed line is the redshift dependent analytic form calibrated by \citet{delhaize17} on a sample of SFGs, with $\pm 1\sigma$ spread (magenta area). White circles show the redshift binned peak values of the IR-to-1.4 GHz luminosity ratio calibrated on the full radio detected sample.}
  \label{fig:q_vs_z_full_sample}
\end{figure}

By studying the sample of 9,575 SFGs in the COSMOS field, \citet{delhaize17} found a redshift dependent IRRC. Their SFG sample contained both direct radio and far-infrared measurements and limits. On the other hand, the full sample considered here contains only radio (3 GHz) detections. For this reason, we derive a separate IRRC by binning this radio detected sample in nine redshift bins (defined as in the analysis of the HLAGN sample). 
For each source we calculate the infrared-to-total 1.4 GHz luminosity ratio ($\mathrm{q_{TOT}}$) as defined in Sec. \ref{sec:decomposition}, Eq.\ref{eq:qtot}. The calculated values are shown in Fig. \ref{fig:q_vs_z_full_sample}. 

To find the median and corresponding 1$\sigma$ confidence range originating from the star-forming processes we locate the peak ($\mathrm{\bar{q}_{SF,z}}$) of the $\mathrm{q_{TOT}}$ distribution in each redshift bin. We then mirror the part of the $\mathrm{q_{TOT}}$ distribution above the location of its peak where we are certain that all radio emission originates from the star formation (e.g. \citealt{bell03}, \citealt{magnelli15}). By fitting a Gaussian functional form to the mirrored distribution, we find the dispersion. To account for the uncertainties on the luminosities used in the calculation of the $\mathrm{q_{TOT}}$, we use a Monte Carlo method. Derived peaks per redshift bin, $\mathrm{\bar{q}_{SF,z}}$, are also shown in Fig. \ref{fig:q_vs_z_full_sample} as white circles.

The IRRC derived in such a way has slightly lower values below $\mathrm{z\sim1.5}$ than the correlation derived by \citet{delhaize17} and increases towards higher redshifts. Observed trend is not in agreement with result by \citeauthor{delhaize17} and most likely occurs due to differently selected samples (e.g. \citealt{sargent10}).

By following the procedure described in Sec. \ref{sec:decomposition}, we derive AGN fractions for all sources within the full sample. For $\mathrm{(77.4\pm1.4)\%}$ of these sources we find radio emission to be dominated by star-forming processes in the radio ($\mathrm{0\leq f_{AGN} \leq 0.5}$), while the rest, $\mathrm{(22.6\pm1.4)\%}$, are dominated by the AGN-related radio emission ($\mathrm{0.5< f_{AGN} \leq 1.0}$).

By imposing a redshift dependent threshold, \citet{delvecchio17} found that $\mathrm{\sim 23\%}$ of the sources in the full sample display radio excess due to the presence of AGN-related radio emission. For a comparison, with the method of statistical decomposition we effectively imposed a `lower' threshold and found evidence of AGN-related emission in $\mathrm{\sim 26\%}$ of the sample. This method enables us to retrieve information on the AGN contribution to the total radio luminosity also in sources which are not dominated by AGN activity in radio. 

\begin{figure*}[!htbp]
  \includegraphics[width=\linewidth]{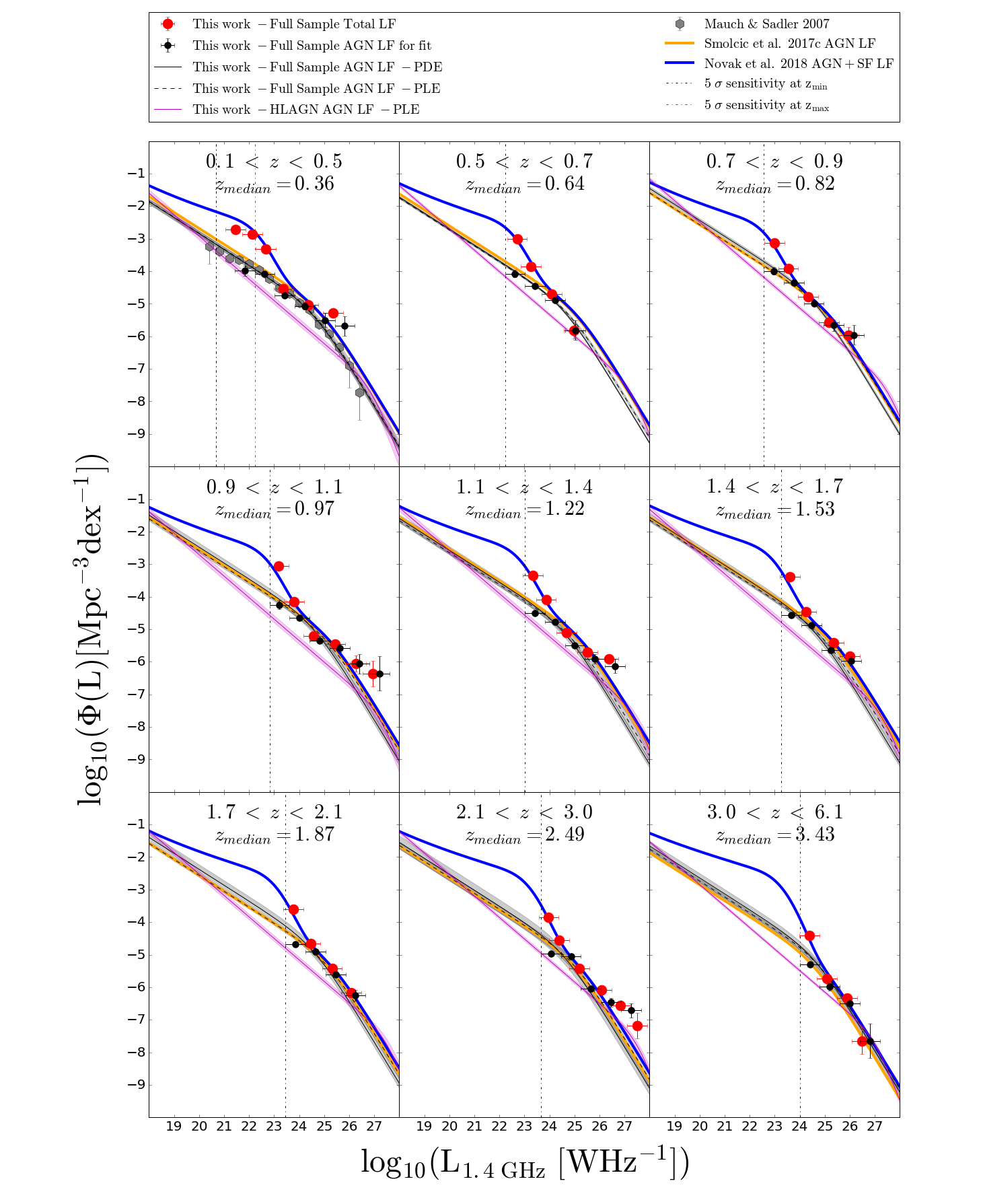}
  \caption{The 1.4 GHz AGN luminosity function of the full VLA-COSMOS 3 GHz Large Project sample with COSMOS2015 counterparts out to $\mathrm{z\sim6}$. The AGN and total luminosity functions are shown with black and red circles, respectively. Solid and dashed black lines show the fit of analytic local AGN luminosity function in PDE and PLE models, respectively, with shaded areas indicating $\mathrm{1\sigma}$ confidence range. The black and red vertical dashed lines show the luminosity corresponding to the 5$\sigma$ sensitivity at $\mathrm{z_{min}}$ and $\mathrm{z_{max}}$ of the redshift bin, respectively. Results from the literature are also shown, as detailed in the legend.}
  \label{fig:lf_combo_total}\vspace{1cm}
\end{figure*}

\subsection{Radio AGN luminosity functions and the cosmic evolution of AGN within the full 3 GHz sample} \label{sec:lf_full_sample}

In this section we construct luminosity functions using the same methods as described in Sec. \ref{sec:luminosity_function}. The total luminosity functions are constructed from the total 1.4 GHz luminosities of all sources within the full sample. The AGN luminosity functions are calculated using the AGN luminosities of sources from the full sample which would be observable by our survey ($\mathrm{S_{3\ GHz, AGN} \geq S_{lim, 3\ GHz}}$). Both the total and AGN luminosity functions of the full sample are shown in Fig. \ref{fig:lf_combo_total}. The values of the AGN luminosity functions are listed in Table \ref{tab:lumfun_full}.

Following the same procedure as in Sec. \ref{sec:lf_evolution_sub}, we use the analytic form of the local luminosity function to trace the cosmic evolution of this sample, which after decomposition contains the entire radio AGN population in the COSMOS field. We use the analytic form of the local luminosity function as derived by \citet{ms07}:
\begin{equation}\label{eq:lf_ms07}
$$\mathrm{\Phi_{0} (L)  = \dfrac{\Phi^{*}}{\left(L^{*}/L\right)^{\alpha} + \left(L^{*}/L\right)^{\beta}},}$$
\end{equation}
with parameters $\mathrm{\Phi^* = 10^{-5.1} \mathrm{Mpc^{-3} dex^{-1}}}$, $\mathrm{L^* = 10^{24.59}\ \mathrm{WHz^{-1}}}$, $\mathrm{\alpha =-1.27}$, $\mathrm{\beta = -0.49}$. 
To constrain the shape of the luminosity function over 6 orders of magnitudes, \citet{ms07} used a sample of 2,661 spectroscopically selected radio loud AGN from a local ($\mathrm{<z>\sim0.073}$) sample of the NVSS sources in the 6 degree Field Galaxy Survey (6dFGS).
We fit the \citeauthor{ms07} analytic form of the local luminosity function to our AGN luminosity functions. Values of evolution parameters derived as described in Sec. \ref{sec:lf_evolution_sub} are shown in Table \ref{tab:alpha_evol_c15} and in Fig. \ref{fig:evol_par_c15}.
Best-fit parameters obtained with a continuous fit of the analytic form are $\mathrm{\alpha_l = 1.97\pm 0.10}$ and $\mathrm{\beta_l=-0.46\pm 0.04}$ in the case of pure luminosity evolution ($\mathrm{\alpha_d=\beta_d =0}$). In the case of pure density evolution ($\mathrm{\alpha_l=\beta_l =0}$) the best-fit parameters are $\mathrm{\alpha_d = 1.24\pm 0.08}$ and $\mathrm{\beta_d =-0.25\pm 0.03}$.

\begin{figure}[!htpb]
  \includegraphics[width=\linewidth]{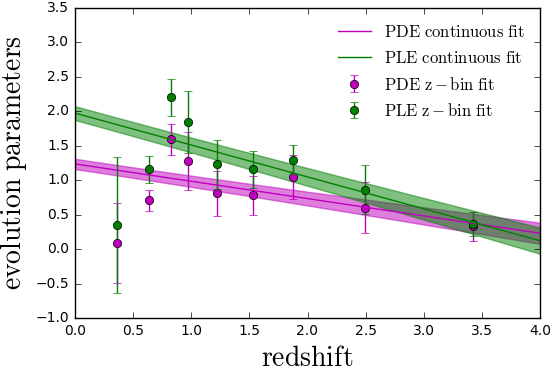}
  \caption{Fitting parameters obtained to model the redshift evolution of the full AGN sample. Green (magenta) circles and line show the pure luminosity (density) evolution parameter values and model, respectively.}
  \label{fig:evol_par_c15}
\end{figure}

\citet{smolcic17c} constructed the 1.4 GHz luminosity functions of radio AGN selected by applying a redshift dependent ($\mathrm{>3\sigma}$) threshold in the ratio of the radio luminosity and SFRs derived from IR luminosities. 
This selection yielded a sample of 1,814 sources in which over $\mathrm{80\%}$ of radio emission comes from an AGN activity. By fitting the analytical form of local luminosity function by \citet{ms07}, as done also here, they constrained the pure density and pure luminosity evolutions of the radio AGN luminosity function out to $\mathrm{z\sim5.5}$. We show their luminosity functions for the PLE model in Fig. \ref{fig:lf_combo_total}. Their results agree well with AGN luminosity functions of the full sample.
The modeling of the redshift evolution of their luminosity function is equivalent to the analysis performed here. The differences found in the evolution parameters in the case of the discrete and continuous fitting of the radio luminosity function come from several factors. To ensure a complete sample, we first impose an AGN flux density cut which reduces the number of sources used in the analysis. We calculate the maximum observable volume of each source using the AGN luminosity derived from the luminosity decomposition analysis, obtaining lower values for the volumes than would be the case if using the total luminosity of the source. Finally, to construct luminosity functions we use the AGN luminosities, which are the total luminosities scaled by AGN fraction. All of these factors result in slightly different values of evolution parameters obtained here and by \citet{smolcic17c}. However, the results are still consistent within $\mathrm{2 \sigma}$.

\begin{table}
\begin{center}
\caption{Best-fit evolution parameters obtained by fitting local luminosity function to redshift binned data assuming pure density ($\mathrm{\alpha_D}$) and pure luminosity ($\mathrm{\alpha_L}$) evolution.}
\renewcommand{\arraystretch}{1.5}
\begin{tabular}[t]{c c c}
\hline
Med(z)
& $\alpha_{D}$
& $\alpha_{L}$\\
\hline 
0.36 & 0.09 $\pm 0.58$ &  0.35 $\pm 0.99$ \\
0.64 & 0.71 $\pm 0.15$ &  1.16 $\pm 0.19$ \\
0.82 & 1.59 $\pm 0.22$ &  2.20 $\pm 0.27$ \\
0.97 & 1.28 $\pm 0.43$ &  1.84 $\pm 0.45$ \\
1.22 & 0.81 $\pm 0.32$ &  1.23 $\pm 0.36$ \\
1.53 & 0.78 $\pm 0.28$ &  1.16 $\pm 0.27$ \\
1.87 & 1.04 $\pm 0.32$ &  1.30 $\pm 0.21$ \\
2.49 & 0.60 $\pm 0.37$ &  0.86 $\pm 0.36$ \\
3.43 & 0.33 $\pm 0.21$ &  0.37 $\pm 0.17$ \\

\hline
\end{tabular}
\quad
\label{tab:alpha_evol_c15}
\end{center}
\end{table}

We also compare our results with the study done by \citet{novak18}. Using a sample of 8,035 radio detected sources from the VLA-COSMOS 3 GHz Large Project with COSMOS2015 (\citealt{laigle16}) and i-band (\citealt{capak07}) multi-wavelength counterparts they constrained the evolution of the total radio luminosity function. To construct the total radio luminosity function, they combine the local luminosity functions of SFGs and of AGN from the literature. Our total (AGN+SF) luminosity functions shown in Fig. \ref{fig:lf_combo_total} are in excellent agreement with their results.

Agreement with the results by \citet{smolcic17c} and \citet{novak18} shows that the high luminosity end of AGN luminosity function as derived here is populated by sources which are predominantly dominated by AGN activity. Sources with lower AGN fractions are more common towards the lower luminosity end of AGN luminosity function, in a regime below the luminosity threshold of our survey. In this regime, the decomposition allows us to go further in determining the true shape of the AGN luminosity function.

Using the evolution parameters as derived with the continuous fit to all data simultaneously, we further calculate the number and luminosity density evolution curves which are described in Sec. \ref{sec:nd_ld} and shown in Fig. \ref{fig:densities}. Using the \citet{willott99} conversion between the monochromatic 1.4 GHz luminosity and the kinetic luminosity, we estimate the kinetic luminosity density of the full sample and discuss it in the context of semi-analytic models from the literature (see Sec. \ref{sec:kinetic_feedback}).

%table for double column format
\begin{table*}
\begin{center}
\caption{AGN luminosity functions for the full VLA-COSMOS 3 GHz Large Project sample with COSMOS2015 multi-wavelength counterparts.}
\renewcommand{\arraystretch}{1.5}
\begin{tabular}[t]{c c c}
\hline
Redshift
& $\log\left(\dfrac{L_{1.4\,\text{GHz}}}{\text{W}\,\text{Hz}^{-1}}\right)$
& $\log\left(\dfrac{\Phi}{\text{Mpc}^{-3}\,\text{dex}^{-1}}\right)$\\
\hline 
$0.10<z<0.50$ & 21.82 &  -3.99 $\pm 0.15$ \\
$Med(z) = 0.36$ & 22.62 &   -4.09 $\pm 0.04$ \\
& 23.42 &  -4.74 $\pm 0.06$ \\
& 24.22 &  -5.08 $\pm 0.10$ \\
& 25.02 &  -5.50 $\pm 0.22$ \\
& 25.82 &  -5.68 $\pm 0.30$ \\
$0.50<z<0.70$ & 22.62 &  -4.07 $\pm 0.11$ \\
$Med(z) = 0.64$ & 23.42 &   -4.45 $\pm 0.05$ \\
& 24.22 &  -4.88 $\pm 0.06$ \\
& 25.02 &  -5.81 $\pm 0.30$ \\
$0.70<z<0.90$ & 22.96 &  -3.99 $\pm 0.11$ \\
$Med(z) = 0.82$ & 23.76 &   -4.35 $\pm 0.09$ \\
& 24.56 &  -4.98 $\pm 0.06$ \\
& 25.36 &  -5.66 $\pm 0.17$ \\
& 26.16 &  -5.96 $\pm 0.30$ \\
$0.90<z<1.10$ & 23.21 &  -4.25 $\pm 0.13$ \\
$Med(z) = 0.97$ & 24.01 &   -4.66 $\pm 0.06$ \\
& 24.81 &  -5.35 $\pm 0.08$ \\
& 25.61 &  -5.58 $\pm 0.12$ \\
& 26.41 &  -6.06 $\pm 0.30$ \\
& 27.21 &  -6.36 $\pm 0.53$ \\

\hline
\end{tabular}
\quad
\renewcommand{\arraystretch}{1.5}
\begin{tabular}[t]{c c c}
\hline
Redshift
& $\log\left(\dfrac{L_{1.4\,\text{GHz}}}{\text{W}\,\text{Hz}^{-1}}\right)$
& $\log\left(\dfrac{\Phi}{\text{Mpc}^{-3}\,\text{dex}^{-1}}\right)$\\
\hline  
$1.10<z<1.40$ &  23.42 &  -4.51 $\pm 0.05$ \\
$Med(z) = 1.22$ & 24.22 &   -4.78 $\pm 0.04$ \\
& 25.02 &  -5.49 $\pm 0.07$ \\
& 25.82 &  -5.91 $\pm 0.14$ \\
& 26.62 &  -6.13 $\pm 0.22$ \\
$1.40<z<1.70$ &  23.66 &  -4.56 $\pm 0.04$ \\
$Med(z) = 1.53$ & 24.46 &   -4.87 $\pm 0.05$ \\
& 25.26 &  -5.64 $\pm 0.08$ \\
& 26.06 &  -5.96 $\pm 0.14$ \\
$1.70<z<2.10$ &  23.85 &  -4.67 $\pm 0.06$ \\
$Med(z) = 1.87$ & 24.65 &   -4.90 $\pm 0.09$ \\
& 25.45 &  -5.62 $\pm 0.06$ \\
& 26.25 &  -6.25 $\pm 0.16$ \\
$2.10<z<3.00$ &  24.06 &  -4.97 $\pm 0.05$ \\
$Med(z) = 2.49$ & 24.86 &   -5.04 $\pm 0.09$ \\
& 25.66 &  -6.03 $\pm 0.08$ \\
& 26.46 &  -6.46 $\pm 0.14$ \\
& 27.26 &  -6.71 $\pm 0.22$ \\
$3.00<z<6.10$ &  24.40 &  -5.30 $\pm 0.07$ \\
$Med(z) = 3.43$ & 25.20 &   -5.99 $\pm 0.11$ \\
& 26.00 &  -6.49 $\pm 0.06$ \\
& 26.80 &  -7.65 $\pm 0.53$ \\

\hline
\end{tabular}
\label{tab:lumfun_full}
\end{center}
\end{table*}

\end{document}